\documentclass[smallextended]{svjour3}       
\smartqed  
\usepackage{graphicx}
\newcommand{\be}{\begin{equation}}
\newcommand{\ee}{\end{equation}}
\newcommand{\bea}{\begin{eqnarray}}
\newcommand{\eea}{\end{eqnarray}}

\begin{document}

\title{Conformal and Affine Hamiltonian Dynamics of General Relativity}

\author{Victor N. Pervushin \and
Andrej B. Arbuzov \and
Boris M. Barbashov \and
Rashid G. Nazmitdinov \and
Andrzej Borowiec \and
Konstantin N. Pichugin \and
Alexander F. Zakharov}

\institute{V.N. Pervushin, A.B. Arbuzov, B.M. Barbashov, R.G. Nazmitdinov, A.F. Zakharov \at
Bogoliubov Laboratory of Theoretical Physics, Joint Institute for Nuclear Research,
141980 Dubna, Russia \\
Tel.: +7-49621-63343 \\
\email{arbuzov@theor.jinr.ru}
\and
A.B. Arbuzov \at  Department of Higher Mathematics,
University of Dubna,  141980 Dubna, Russia
\and
R.G. Nazmitdinov \at 
Department de F{\'\i}sica,
Universitat de les Illes Balears, E-07122 Palma de Mallorca, Spain
\and
A. Borowiec \at
Institute of Theoretical Physics, University of
 Wroc\l aw, Pl. Maxa Borna 9, 50-204 Wroc\l aw, Poland
\and
K.N. Pichugin \at
Kirensky Institute of Physics, 660036 Krasnoyarsk,  Russia
\and 
A.F. Zakharov \at
Institute of Theoretical and Experimental Physics,
B. Cheremushkinskaya str. 25, 117259 Moscow, Russia}


\date{Received: date / Accepted: date}

\maketitle

\begin{abstract}
The Hamiltonian approach to the General Relativity is formulated as a joint
nonlinear realization of conformal and affine symmetries by means of  the Dirac scalar
dilaton and  the Maurer-Cartan forms. The dominance of the Casimir
vacuum energy of physical fields provides a good description of the type
Ia supernova luminosity distance--redshift relation. Introducing the uncertainty
principle at the Planck's epoch within our model, we obtain the hierarchy of the
Universe energy scales, which is supported by the observational data.
We found that the  invariance of the Maurer-Cartan forms with respect to the general
coordinate transformation yields a single-component strong gravitational waves.
The Hamiltonian dynamics of the model describes the effect of an intensive vacuum
creation of gravitons and  the minimal coupling scalar (Higgs)
bosons in the Early Universe.
\end{abstract}

\section{Introduction
\label{sect_Intro}}

The Standard Model (SM) of electroweak and strong
interactions~\cite{weak-67} describes well practically all physical phenomena
up to the energy of at least 100~GeV. According to the accepted wisdom, the
physical content of the SM (in the lowest order of perturbation theory) is
completely covered  by fields and particles as representations of the
Poincar\'e group (with positively defined Hilbert space scalar product) and
the Lorentz classification of variables used for calculation of weak
transitions between these states~\cite{logunov}.

The unification of the SM with a gravitational theory is a longstanding
fundamental problem.
It seems natural in this case to require that both theories
should be treated on equal footing. The main difficulty of the unification
lies in the different theoretical levels of their presentation: quantum for
the SM and classical for the gravitational theory. However, both these
theories have common roots of their origin (mechanics and electrodynamics) and
obey the \emph{principles of relativity} confirmed by numerous experimental
observations. Note that there is also enormous progress in observational
cosmology~\cite{Giovannini,rub-09} which enters into the era of precise
science; it means that a typical accuracy of standard parameter determination
is about a few per cent. Evidently, one of the major aims of the unification
is to develop a cosmological model which could pass the vitality test by the
cosmological data. Last but not least, this model should allow one to develop
a renormalizable quantized version.

The first step in  this direction is due to Fock ~\cite{Fock:1929vt}. He introduced
the Dirac electron as a spinor representation of the Lorentz group into the
General Relativity (GR) by means  of the Einstein  interval as the sum of
squared linear forms. These forms are known as four components of  a simplex
of the reference frame in the tangent Minkowskian space-time. The next step
was made in~\cite{og-73}, where  it was shown that infinite-parametric general
coordinate transformations introduced by Einstein~\cite{Einstein} can be
converted to the finite-parametric conformal group and the affine group of all
linear transformations of four-dimensional space-time, including the
Poincar\'e group.
The  conformal symmetry  as  a basis for the
 construction of the GR was independently introduced
by Deser and Dirac~\cite{Deser:1970hs,Dirac_73}. In particular, Dirac
formulated the conformal-invariant approach to the GR~\cite{Dirac_73} as a new
variational principle for the Hilbert action~\cite{H_1915}
 introducing a dilaton (scalar) field, in
addition to $g_{\mu\nu}$. Further it was shown~\cite{og-74} that in the case
of the dynamical affine symmetries the method of nonlinear realization of
symmetry
groups~\cite{Col_69} leads to the Hilbert action
of Einstein's gravitational theory expressed  in terms of the Fock simplex
components.

A particular conformal cosmological model, based on the ideas discussed above,
have been developed in
papers~\cite{Behnke:2001nw,Zakharov:2010nf,Barbashov:2005hu,Arbuzov:2010fz}.
It was shown that the model leads
to a viable cosmology being in agreement with observations. For example,
a good description of the modern supernovae type Ia data was
constructed~\cite{Behnke:2001nw,Zakharov:2010nf}
in the assumption of a rigid state dominance.
The initial data conditions on the dilaton and
an additional scalar field have been employed
as a source of the conformal symmetry breaking.
This field led to the rigid state dominance at the classical level.
In the present paper we shall attempt to go beyond the
classical level and show that the Casimir vacuum effect in a finite-size
Universe could provide both the scale invariance breaking and the rigid state
dominance, required in our model to describe the SNe~Ia data.
Therefore we substantially change the basis of the whole model.
For this reason we reformulated below the model and re-derived its
phenomenological consequences.
Note that recently different approaches to construction of conformal-invariant
models of gravity and cosmology have been suggested in the literature,
see {\it e.g.} papers~\cite{Blas:2011ac,GarciaBellido:2011de,Penrose:1972tt} and
references therein.

The  goal objective of our paper is to construct a self-consistent
gravitational model of the Universe based on the
affine and conformal symmetries
in the framework of the Dirac variational principle.
Our approach enables us: i) to develop the Hamiltonian description of the cosmological
evolution, ii) to obtain exact solution of constraints, and iii) using this
solution to  gain cosmological quantum effects, including the vacuum creation
processes at the Planck epoch.
The content of the paper is as follows. Section~\ref{sect_HD} is devoted to
the Hamiltonian approach to our gravitation model, clearing up  its symplectic
structure. Here we  also establish the Planck epoch hierarchy of
the Universe energy scales defined by  the Casimir vacuum energy and  the
quantum uncertainty principle. In Section~\ref{sect_CC}, we
analyze the  properties of strong affine gravitational waves
in a dynamical approximation, when all static Newton-like potentials are
neglected. The conformal cosmological (CC) model is briefly discussed and
an intensive vacuum creation of gravitons is described.
Section~\ref{sect_matter} describes gravitational interactions of fermions.
In Section~\ref{sect_Higgs} we  compare the vacuum creation of gravitons with the
Higgs particle one.  The main results are summarized in
Section~\ref{sect_Summ}.  In Appendix~A, we briefly reconsider the Dirac
Hamiltonian approach in infinite space-time, reformulated in terms of the
Maurer--Cartan  forms in order to compare it with our construction in Section
\ref{sect_HD}. In Appendix~B we present the dilaton cosmological perturbation
theory. Appendix~C is devoted to the conformal cosmology.

\section{Conformal Hamiltonian dynamics
\label{sect_HD}}

\subsection{Nonlinear realization of affine and conformal symmetries}

Let us define a conformal version of the GR as a nonlinear realization of
joint conformal and affine $A(4)$
symmetries  in the factor space $A(4)/L$ with the Lorentz subgroup $L$
of the stable vacuum (here we use the concepts of the theory~\cite{Col_69}).
Recall that the affine group $A(4)$ is the group of all linear
transformations of the four-dimensional manifold $x^\mu\to$
$\widetilde{x}^\mu={x}^\mu+y^\mu+L_{[\mu\nu]}{x}^\nu+R_{\{\mu\nu\}}{x}^\nu$,
where $y^\mu$ is a  shift of coordinate and $L_{[\mu\nu]}$ and $R_{\{\mu\nu\}}$
are antisymmetric and symmetric matrices respectively
(here Greek indices $\mu,\nu,\ldots$ run from 0 to 3).
A nonlinear realization of $A(4)$ is based on
finite transformations $G=e^{iP\cdot x}e^{iR\cdot h}$
defined by means of the
shift operator $P$, proper affine transformation $R$ and
the following  Goldstone modes:
four coordinates $x_{\mu}$  and ten gravitational fields $h$~\cite{og-73}.
This realization
can be  obtained with the aid of the Maurer--Cartan forms\footnote{These forms were
introduced in the GR by   Fock and Cartan~\cite{Fock:1929vt,car}).}
in the following way
 \bea \label{1-0}
 GdG^{-1}&=&i[\underbrace{P_{(\alpha)}\cdot
 \omega^P_{(\alpha)}+R_{(\alpha)(\beta)}\cdot
 \omega^R_{(\alpha)(\beta)}}_{{\rm shifts~of~simplex~in}~K=A(4)/L}
 +\underbrace{L_{(\alpha)(\beta)}\cdot
 \omega^L_{(\alpha)(\beta)}}_{{\rm rotations~in}~K=A(4)/L}],
 \\ \label{pR}
 \omega^P_{(\alpha)}(d)&=&e_{(\alpha)\mu}dx^\mu,
 \\ \label{tR}
 \omega^R_{(\alpha)(\beta)}(d) &=& \frac{1}{2}
 \left(e^{\mu}_{ (\alpha)}de_{(\beta)\mu }
 +e^{\mu}_{(\beta)}de_{(\alpha)\mu }\right),
 \\ \label{tL}
 \omega^L_{(\alpha)(\beta)} (d) &=& \frac{1}{2}
 \left(e^{\mu}_{ (\alpha)}de_{(\beta)\mu }
 -e^{\mu}_{ (\beta)}de_{(\alpha)\mu }\right).
 \eea
 Here,  $\omega^P_{(\alpha)}(d)$, $\omega^R_{(\alpha)(\beta)}(d)$ are
shifts of a simplex of the reference frame
in  the coset space $A(4)/L$, and  $\omega^L_{(\alpha)(\beta)}(d)$
is responsible for the rotation of the simplex.
The explicit dependence of the decomposition coefficient
$e^{\mu}_{(\alpha)}, e_{(\beta)\mu }$ (known as tetrades~\cite{Fock:1929vt})
on the gravitational fields
$h$  was obtained in Refs.~\cite{og-74,per76}.
Note that there are two types of indices: one belongs to the
subgroup $L$  and the other (bracket Greek indices $(\alpha),(\beta),\ldots$ run
from 0 to 3)
to the coset $A(4)/L$.
In this approach, the Maurer--Cartan forms with the coset indices are
main objects of the Poincar\'e  transformations and classification of states.
According to the general theory of non-linear realizations~\cite{Col_69}
we should express all measurable quantities via the coset variables with
bracket-indices.

To construct a GR model in this approach, one needs to
consider the covariant differentiation of a set of fields
$\Psi$ transformed by means of
 the Lorentz group generators $L^{\Psi}_{(\alpha)(\beta)}$
 \bea\label{fock}
D_{(\gamma)}\Psi&=&\frac{D\Psi}{\omega^P_{(\gamma)}}=
 \left[\partial_{(\gamma)}
 +\frac{1}{2}i
v_{(\alpha) (\beta) ,(\gamma)} L^{\Psi}_{(\alpha) (\beta) }\right]\Psi,
 \eea
where $\partial_{(\gamma)}=(e^{-1})_{ \mu
(\gamma)}\partial_{\mu}$.
Here, the linear form $v_{(\alpha) (\beta),(\gamma)}$
is constructed by means of the Maurer--Cartan forms (\ref{tR})
(\ref{tL}) ,
 \bea
\label{carta1}
v_{(\alpha) (\beta),(\gamma)}= \left[\omega^L_{(\alpha)(
\beta)}(\partial_{(\gamma)})+ \omega^R_{(\alpha)(\gamma)}(\partial_{(\beta)})
-\omega^R_{(\beta )(\gamma)}(\partial_{(\alpha)})\right].
 \eea
Similarly, the covariant expression
for the action of the Goldstone fields $h$ can be obtained with the
aid of the commutator of the covariant differentiation of a set of
the  fields $\Psi$~\cite{ll}
 \be \left[
 D_{(\delta)}D_{(\gamma)}-D_{(\gamma)}D_{(\delta)}\right]\Psi
 =iR^{(4)}_{(\alpha)(\beta),(\delta)(\gamma)} L^{\Psi}_{(\alpha) (\beta)}
 \Psi/2.
 \ee
 The  Riemann curvature tensor is defined as~\cite{og-74}:
 \bea
\label{curv}
&& R^{(4)}_{(\alpha)(\beta),(\gamma)(\delta)} =
\partial_{(\gamma)} \; v_{(\alpha) (\beta),(\delta)}+
v_{(\alpha) (\beta) ,(\zeta)} \; v_{(\delta) (\zeta),(\gamma)}
\nonumber\\
&& \quad + v_{(\alpha)(\zeta),(\delta)} \; v_{(\beta) (\zeta),(\gamma)}-((\gamma)
~\leftrightarrow~(\delta))
 \eea
The joint conformal realization
of the affine and conformal groups $A(4)\times C$ symmetry
allows to separate the dilaton field D~\cite{Dirac_73}  as the Goldstone mode
accompanying the spontaneous conformal
symmetry breaking via a scale transformation:
\bea
\label{Gold-1}
e^\mu_{(\alpha)}=\widetilde{e}^\mu_{(\alpha)} e^D.
 \eea

Using the correspondence principle to the classical gravitation theory
and the minimal derivative number
postulates, we obtain the conformal-invariant action:
 \bea
 \label{1-1}
 W_{\rm C}[D,\widetilde{e}_{(\alpha)\nu}] &=&
 - M^2_{\rm C}\frac{3}{8\pi}
 \int\limits_{ }^{ }\! d^4x\,
 \biggl[
  \frac{\sqrt{-\widetilde{g}}}{6}\,{R}^{(4)}(\widetilde{g})
 \,e^{-2D} 
\nonumber \\
&-& e^{-D}\,\partial_\mu
\, \left(\sqrt{-\widetilde{g}}\,\widetilde{g}^{\mu\nu}\,\partial_\nu e^{-D}\,
\right) \biggr],
 \eea
 where $M_C$ is the conformal Newton coupling constant.

 In this case the measurable interval $\widetilde{ds}^2$
 is determined by the conformal metric $\widetilde{g}_{\mu\nu}$
 expressed via the tetrades $\widetilde{e}_{(\alpha)\mu}$:
 \bea\label{PVmain}
 {\widetilde{g}}_{\mu\nu}=\widetilde{e}_{(\alpha)\mu}\otimes\widetilde{e}_{(\alpha)\nu}\
\to \
 \widetilde{ds}^2={\widetilde{g}}_{\mu\nu}dx^\mu dx^\nu.
 \eea

Note that the standard Hilbert--Einstein action
 \bea\label{D-2gca}
 W_{\rm E}[g]&=&-(M_{\rm Pl}^2/16)\int d^4x\sqrt{-g}R^{(4)}(g)
 \eea
 with the standard Einstein interval
 \bea
 \label{D-2gc}
 {ds}_{\rm E}^2&=&{g}_{\mu\nu}dx^\mu dx^\nu
 \eea
can be reproduced:
 \bea
 \label{WC_WE}
 W_{\rm C}[D,\widetilde{e}_{(\alpha)\nu}]= W_{\rm E}[g], \qquad {\mathrm{if}} \qquad
 \left\{ \begin{array}{ll}
  g_{\mu\nu} =
 e^{2D}\widetilde{e}_{(\alpha)\nu} \otimes\widetilde{e}_{(\alpha)\nu}\\
 M_{\rm Pl} = M_{\rm C}
 \end{array}
  \right.
  \eea

In the action~(\ref{1-1}) we have joined two approaches:
the Dirac's dilaton conformal theory and the Fock tetrades with the affine symmetry.
Although there is a formal correspondence (\ref{WC_WE}),
some physical consequences, as will be shown
below, are different. We point out that in our approach there is a new set of
dynamical variables $\{\widetilde{e}^\mu_{(\alpha)},D\}$ which are subject to
the affine and conformal symmetry constraints. In particular, the conformal
invariant interval $\widetilde{ds}^2$ substitutes the standard one ${ds}_{\rm E}^2$.
It will be shown below that the new variables enable to us  to find explicit solutions for
all symmetry constraints in the framework of the Dirac approach~\cite{dir}.
One of the key  assumptions of our approach to the GR is
that  measurable quantities are identified with the conformal field variables
$\widetilde{F}_{(n)}$. These variables are obtained from the standard ones ${F}_{(n)}$
by means of the Weyl transformation
${F}_{(n)}=\widetilde{F}_{(n)}e^{nD}$, where $n$ is the conformal
weight~\cite{Weyl_18}. Therefore, we name our approach as a conformal general
relativity (CGR) approach to the gravitation theory.

Below we will use the {\it natural units}:
 \be
 \label{units1} M_{\rm Pl}\sqrt{3/(8\pi)}=c=\hbar=1.
 \ee

\subsection{Conformal formulation of the Dirac-ADM foliation $4\to 1+3$}

Thus, we have defined the action and the variables of our model.
In order to obtain physical results we have to resolve within the Hamiltonian
approach the constraints arisen due to the affine and conformal symmetries.

Let us reformulate the Dirac-ADM foliation~\cite{dir,ADM} in terms of the simplex
components
and the dilaton\footnote{
Although the Dirac Hamiltonian approach to the Hilbert action in terms of the
Dirac-ADM metric components is well known~\cite{dir,ADM}, for the sake of comparison
we present in Appendix A the modification of this approach in terms of the
Maurer--Cartan forms.}.
The simplex components
$[{\widetilde{\omega}}_{(0)},{\widetilde{\omega}}_{(b)}]$
(here all Latin indices run from 1 to 3)
can be written as
 \bea \label{dg-2aa}
 {\widetilde{\omega}}_{(0)}&=&e^{-2{D}}N dx^0,
 \\ \label{dg-2aa1}
 {\widetilde{\omega}}_{(b)}&=&\widetilde{{\bf e}}_{(b)i}dx^i+{N}_{(b)}dx^0,
 \eea
where
${N}_{(b)}=N^j\widetilde{{\bf e}}_{j(b)}$ are the shift vector
components, and  $N(x^0,x^j)$ is the lapse function.
Here $\widetilde{\omega}_{(b)}$ are the linear
forms defined via the triads $\widetilde{{\bf e}}_{(b)i}$ with a unit spatial
metric determinant \be\label{frame-1}
 |\widetilde{{\bf e}}_{j(b)}|=1\;,
\ee
i.e., the  Lichnerowicz gauge~\cite{lich}-type for the triads.
This gauge connects the scalar
dilaton field $D$ with a logarithm of the Einstein metric determinant:
 \be\label{D-1d}
 D= -(1/6)\ln|{g}^{(3)}_{ij}|.
 \ee
Recall that this component was distinguished by Dirac in his Hamiltonian approach to
the Einstein GR~\cite{dir}.

The group of invariance of the GR for  the Dirac-ADM foliation is known
as the kinemetric subgroup of the general coordinate transformation~\cite{vlad}
  \bea
  \label{1-1aa}
  x^0&\to& \widetilde{x}^0=\widetilde{x}^0(x^0),\\
  \label{1-1b}
 x^k&\to& \widetilde{x}^k=\widetilde{x}^k(x^0,x^1,x^2,x^3).
 \eea
This group admits the
decomposition of the dilaton  into the sum of the zeroth and nonzeroth harmonics:
 \be\label{1-1dd}
 D(x^0,x^1,x^2,x^3)=\langle D\rangle(x^0)+ \overline{D}(x^0,x^1,x^2,x^3).
 \ee
 This is one of the key points in our construction.
The introduction of the zeroth mode $\langle D\rangle(x^0)$ is consistent with
the Einstein cosmological principle of averaging of all scalar fields of the
theory over a finite volume $V_0=\int_{V_0} {d^3x}$~\cite{Einstein:1917ce}
 \bea
  \label{1-avd}
 &&\langle D \rangle(x^0)=V_0^{-1}\int_{V_0} {d^3x}
 D(x^0,x^1,x^2,x^3).
\eea
In virtue of Eqs.~(\ref{1-1dd}) and (\ref{1-avd}), we obtain the orthogonality condition
\be
\label{1-ortd}
 \int_{V_0} {d^3x} \overline{D}(x^0,x^1,x^2,x^3)\equiv 0.
\ee
 This condition enables to us to consider the zeroth and nonzeroth
 components as independent ones.

The invariance of the action with respect to the reparameterization of the
coordinate time parameter (\ref{1-1aa}) guides us to suppose that
the zeroth dilaton mode $\langle D\rangle(x^0)$ can be chosen
    as an evolution parameter
in the field space of events
 $[\langle D\rangle,\overline{D},{\bf e}^j_{(b)}]$~\cite{Barbashov:2005hu}.
Note that by definition the zeroth dilaton harmonics (obtained by averaging it  over a
finite volume) coincides with the cosmological scale factor logarithm~\cite{Friedman:1922kd}
 \bea\label{1-0bb}
\langle D \rangle=-\ln a=\ln(1+z).
 \eea

The factorization of the lapse function
\bea
\label{1-avd1f}
 N(x^0,x^j)=N_0(x^0){\cal N}(x^0,x^j)
\eea
 by the spatial volume average~\cite{Barbashov:2005hu}
\bea
\label{1-avd11}
 \langle N^{-1}\rangle\equiv \frac{1}{V_0}\int_{V_0} {d^3x}
 \frac{1}{N(x^0,x^1,x^2,x^3)}=N^{-1}_0(x^0).
\eea
yields
the diffeo-invariant proper dilaton time
interval $d\tau$  connected with the world
time interval $dt$ and the conformal one $d\eta$ as
 \be
 \label{eta-1}
  d\tau=N_0(x^0)dx^0=a^{-2}d\eta=a^{-3}dt.
 \ee
 Here, we used the obvious normalization condition for the diffeo-invariant
 lapse function
 \be\label{eta-m1}
 \langle {\cal N}^{-1}\rangle\equiv\frac{1}{V_0}\int_{V_0} {d^3x}
 \frac{1}{{\cal N}(x^0,x^j)}=1.
 \ee
This classification of time-intervals (dilaton, conformal,
and world ones) enables one to introduce the corresponding Hubble parameters
 \bea\label{hub-tau}
 H_\tau &\equiv& -\partial_\tau \langle D \rangle,
 \\\label{hub-eta}
 H_\eta &\equiv& -\partial_\eta \langle D \rangle,
 \\\label{hub-t}
 H_t &\equiv& -\partial_t \langle D \rangle.
\eea

The choice of
the zeroth dilaton mode $\langle D\rangle$   as an evolution parameter
has two consequences in the Hamiltonian approach.
First,  the zeroth dilaton mode canonical momentum density
 \bea \label{d-33}
 P_{\langle D \rangle}&=&
 \frac{2}{V_0}\int_{V_0}d^3x \sqrt{-g}g^{00}\frac{d}{dx^0} \langle D \rangle
  \equiv 2 \frac{d}{d\tau}\langle D \rangle =2 v_{\langle D\rangle}={\rm Const.}\neq 0
\eea
can be treated as a generator of the Hamiltonian evolution in the
 field space of events~\cite{DeWitt:1967yk,WDW}. Here $v_{\langle D\rangle}$ is the
corresponding
 velocity, by construction it coincides with $H_\tau $ introduced in
Eq.~(\ref{hub-tau}).
 We stress that the scale-invariance ($D\to D+\Omega$) admits only a constant
$P_{\langle D \rangle}$.

The second consequence of the  orthogonality condition (\ref{1-ortd})
is that the nonzeroth
harmonics $\overline{D}(x^0,x^1,x^2,x^3)$ do not depend on the evolution parameter.
Therefore, the canonical momentum of dilaton nonzeroth modes is equal to zero:
 \be
 \label{D-2a}
 P_{{\overline{D}}}/2= v_{\overline{D}}= \left[(\partial_0-N^l\partial_l){\overline{D}}
 +\partial_lN^l/3\right]/N = 0.
 \ee
Note that in the Dirac approach, the condition $v_{\overline{D}}=0$ was
also introduced as an additional second class constraint~\cite{dir,Faddeev:1973zb},
see Appendix~A.
In this case the nonzeroth modes play the role of gravitational Newton-type potentials
as the lapse function and the shift vector do.

This result fixes the longitudinal shift vector component (\ref{dv-8}).
As a result, we have
 \be\label{ort-1v}
 \int d^3x v_{\langle D\rangle}\cdot v_{\overline{D}}=0\;,
 \ee
that follows from of Eqs.~(\ref{1-1dd}), (\ref{1-avd}), and (\ref{1-ortd}).
The  orthogonality conditions  (\ref{1-ortd}) and (\ref{ort-1v}) preserve
the definite metrics in the Hilbert space of states~\cite{logunov,grib80}.

Thus taking into account Eqs.~(\ref{1-1dd}) and (\ref{1-0bb})),
we have the following action:
 \bea
 \label{1-3n}
 &&W_{\rm C}\ =\ \underbrace{W_{\rm
 Universe}}_{\rm =0~for~V_0= \infty} \ + \ W_{\rm graviton} \ + \ W_{\rm potential},
 \eea
 \bea
 \label{1-3nt}
 && W_{\rm Universe}=
 -V_0\int\limits_{\tau_I}^{\tau_0} \underbrace{dx^0 N_0}_{=d\tau}
 \left[\left(\frac{d\, \langle D \rangle}{N_0dx^0}\right)^2
 +\rho_\tau^{\rm v}\right],
\eea
\bea
 \label{1-3ng}
 && W_{\rm graviton}=\int\limits_{}^{}
 d^4x\frac{N}{6}\left[{v_{(a)(b)}v_{(a)(b)}}-e^{-4D}R^{(3)}(\widetilde{{\bf
e}})\right],
  \\\label{1-3np}
 && W_{\rm potential}=\int\limits_{}^{}
 d^4x{N}\left[
 \underbrace{\frac{4}{3}e^{-7D/2}\triangle^{(3)} e^{-D/2}}_{\rm Newtonian~potentials
 }
 \right],
 \eea
 where all definitions  are given in the Appendix A devoted
  to the  Dirac Hamiltonian approach to the GR in terms of the tetrades
 (see Eqs.~(\ref{1-1dc}) -- (\ref{1-20})).
In particular, $N_0$ is the collective lapse function (\ref{1-avd11}),
$v_{(a)(b)}$ are given by Eq.(\ref{1-3-dvvc}),
and the three-dimensional curvature $R^{(3)}(\widetilde{{\bf e}})$ is defined by
Eq.~(\ref{1-17a}).

The action~(\ref{1-3n}) and its representation with the aid of  two last
terms $W_{\rm graviton}$ and $W_{\rm potential}$ is well known.
We just reformulated the action in terms of
conformal and affine variables (given below in a definite Dirac-ADM frame: 4
$\to$ 3+1
(\ref{dg-2aa}) and  (\ref{dg-2aa1}) and 3 $\to$ 2+1 (\ref{dg-2a2}), (\ref{dg-2a3}),
and (\ref{dg-2ax})).
The term $W_{\rm Universe}$ was suggested in Ref.~\cite{Barbashov:2005hu} due to
the separation of the dilaton zeroth mode. Here we introduce a new term, $\rho_\tau^{\rm v}$
as a vacuum graviton energy contribution
(and other contributions from fields if they are taken into account).
The effect of this new term will be discussed below. In Appendix~C
we show also that this term allows to obtain a good description of supernovae data
developed earlier in the conformal cosmological
model~\cite{Behnke:2001nw,Zakharov:2010nf}. In the later one the contribution of
the auxiliary scalar field was exploited instead of $\rho_\tau^{\rm v}$.
Strong gravitational waves within our model will be discussed in Section~3.

The introduction of the finite volume $V_0=\int_{V_0} {d^3x}<\infty$ in $W_{\rm Universe}$
creates a dimensional parameter, and therefore, it breaks  the conformal symmetry.
According to the general wisdom~\cite{Col_69}, the symmetry breaking leads to appearance
of a Goldstone mode~\cite{grib80,G-61}.
It is just the zeroth harmonic $\langle D \rangle$.
Note, however, that the Hamiltonian dynamics governed by the equations of motion
must obey the conformal symmetry (see below).

Thus, the action~(\ref{1-3n}), complemented by the condition (\ref{D-2a})
and field-space evolution generator (\ref{d-33}), provides the framework of the Hamiltonian dynamics
in terms of the variables (\ref{dg-2aa}), (\ref{dg-2aa1}).
This dynamics enables one to determine the
perturbation series ${\cal N}=1+\overline{\delta}\ldots$ with
the consistent constraint  $\int d^3x\overline{\delta}=0$
in the frame of reference co-moving with the local volume element
according to the constraint (\ref{D-2a}) (see Appendix B).

\subsection{The empty Universe limit}
\label{SEU}

At the beginning of Universe, in the limit of the tremendous
values of the $z$-factor ($a\rightarrow 0$),
the action $W_{\rm Universe}$ dominates.
Therefore, it is natural to neglect the last two terms in Eq.~(\ref{1-3n}), i.e.,
we consider an empty space.

Recall that in our approach there are two independent variables:
the dilaton zeroth mode $\langle D \rangle$ and
the global lapse function $N_0$.
The variation of action (\ref{1-3nt})
with respect to the dilaton zeroth mode
leads to  the equation of motion:
\bea
\label{3-10a}
 \frac{\delta W_{\rm Universe}}{\delta\langle D \rangle}=0\Rightarrow
 2\partial_{\tau}\left[\partial_{\tau} \langle D \rangle\right]
 =\frac{d\rho_\tau^{\rm v}}{d\langle D \rangle}\,.
\eea

The variation with respect to the global lapse function
leads to the energy constraint
 \bea
 \label{w-6if}
 \frac{\delta W_{\rm Universe}}{\delta N_0}=0 \Rightarrow
 \left[\partial_{\tau} \langle D \rangle\right]^2=\rho_\tau^{\rm v}.
 \eea
This constraint preserves the conformal symmetry of equation of motion (\ref{3-10a})
with respect to transformations $\langle D \rangle \to \langle D \rangle+C$, if
\bea
 \label{w-6ifcc}
\rho_\tau^{\rm v}
\equiv H_\tau^2=H_0^2
 =\mbox{\rm Const}.
 \eea
 The solution of Eqs.~(\ref{w-6if}), (\ref{w-6ifcc}),
determines Eq.(\ref{1-0bb}) in terms of the dilaton time
 \be
 \label{ec-3}
  \langle D\rangle\equiv\ln(1+z)\equiv-\ln a=H_0\tau\;,
  \ee
which describes the evolution of the redshift with respect to the dilaton time
interval $d\tau$.

Note that our equations  (\ref{3-10a}), (\ref{w-6if}) do not differ from
the original Friedmann's ones written in terms of conformal coordinates
and observable quantities for a rigid state.
Indeed, taking into account Eqs.~(\ref{eta-1}), (\ref{ec-3}), one finds that
Eq.(\ref{w-6if})  has the rigid state
form in terms of the conformal variables (see also Appendix C)
\be
\label{eta-2}
[\partial_\eta a]^2=\rho_{\rm cr}/{a^2}\;,
\ee
where
\be \label{cr-1a}
  \rho_{\rm cr}=H^2_0 \frac{3M^2_{\rm Pl}}{8\pi}=H^2_0
 \ee
is the critical density. This equation leads  to the definition
of the rigid state horizon
 \bea
 \label{1-3hh}
 d_{\rm hor}({a})=
 2 \int\limits_{a_I\to 0}^{{a}}
 d \overline{a}\;
 \frac{\overline{a}}{\sqrt{\rho_{\rm cr}}}=\frac{a^2}{H_0}\;.
\eea
 The evolution of the cosmological scale factor in terms of the conformal time-interval
  given by Eqs.~(\ref{eta-1}), (\ref{eta-2})  yields the
 coordinate distance -- redshift relation for the photon at its light cone
$ds^2_{\rm C}=d\eta^2-dr^2=0$
\be\label{ec-4}
e^{-\langle D\rangle}=a(\eta)=\sqrt{1+2H_0(\eta-\eta_0)};~~~~~~r=\eta-\eta_0,
\ee
as the solution of Eqs. (\ref{w-6if}) and (\ref{w-6ifcc}) in terms of the
conformal variables (\ref{eta-1}).
It coincides with the Friedmann solution of his equation (C.2) for the dominant
rigid state. Here $\eta$ is the instant of the
photon emission by a cosmic atom and
$\eta_0$ is the time of the photon  detection at the Earth.
 In the CGR, the  cosmological scale factor (\ref{ec-4}) provides the
cosmic evolution of atomic masses $m(\eta)=a(\eta)m_0$
which gives the redshift of the cosmic atom spectrum lines: the far is an atom, the
more  is
its redshift.
Therefore, the redshift is produced by the ratio
$\lambda(\eta)/\lambda(\eta_0)$, where $\lambda(\eta)$ is
the photon wave length of the photon emitted by cosmic atom with the mass
$m(\eta)=a(\eta)m_0$ and detected at the Earth, where
an etalon atom at the Earth has the mass $m_0$.

If a measurable photon time is identified with the conformal one,
the square root of the conformal time in Eq.(\ref{ec-4}) means that the Universe was
in the $1/a^2$ regime (\ref{eta-2})
in the epoch of the chemical evolution.
The estimation of the primordial helium abundance~\cite{Fukugita:1997bi,Behnke_04}
takes into account the square root dependence of the $z$-factor
on the measurable time-interval
$(1+z)^{-1}\sim \sqrt{\rm t_{\rm measurable}}$.
In the standard cosmology, where the measurable time-interval is identified with
the Friedmann time, this square root dependence of the $z$-factor is explained by the
radiation dominance. In the conformal cosmology, where the measurable time-interval
is identified with
 the conformal time, the square root dependence of the $z$-factor
is  explained by the universal rigid state $(1+z)^{-1}=
a_I\sqrt{1+2H_I(\eta-\eta_I)}$~\cite{Behnke_04}.

Thus, we found that the empty Universe evolves in time
as a rigid state. Below we demonstrate that the same
$1/a^2$ dependence is also a feature of the Casimir vacuum energy.

\subsection{Conformal Casimir energy and the Universe horizon}

Let us again consider the Early Universe. We assume that
at the instant of creation the world was empty and finite in size.
Therefore, its energy can be associated with the quantum
Casimir energy of all physical fields in the given space.
We shall treat all these fields as massless since
$m(a)\stackrel{a\to0}{\longrightarrow}0$ in the Early Universe epoch.

The Casimir energy of a massless field $f$
\be\label{cas-1}
\textsf{H}_{\rm Cas}^{(f)}=
   \sum\limits_{\bf k}\frac{\sqrt{{\bf k}}^2}{2}=
   \frac{\widetilde{\gamma}^{(f)}}{d_{\rm Cas}(a)}.
\ee
 depends on the geometry, size $d_{\rm Cas}$, topology, boundary conditions, and spin
(in particular, for a sphere of diameter $d_{\rm Cas}$ the number of
$\widetilde{\gamma}\sim 0.1 \div 0.03$) ~\cite{grib80,sh-79}. For simplicity
we assume that the Universe has a spherical volume limited by the horizon.

It is natural to suggest that the energy of a massless field is  proportional
to the inverse visual size of the Universe $d_{\rm Cas}({a})$.
Assuming the same dependence for all fields,
we define the total Casimir energy density of the Universe
summing over all fields
\be\label{cas-1f}
 \rho_\eta^{\rm v}(a)=\sum_f\frac{\textsf{H}_{\rm Cas}^{(f)}}{V_0}=
 \frac{C_0}{d_{\rm Cas}(a)}.
\ee The key assumption of our model is that the Casimir dimension $d_{\rm
Cas}({a})$ is equal to the Universe visual size (its horizon (\ref{1-3hh1}))
 \be
 \label{cas-3}
 d_{\rm Cas}({a})\equiv d_{\rm hor}(a)=2
 \int\limits_{a_I\to 0}^{{a}}
 d \overline{a}\;\;[\rho_\eta^{\rm v}(\overline{a})]^{-1/2}
={2C_0}^{-1/2}\int\limits_{a_I\to 0}^{{a}}
 d \overline{a}\;\;d_{\rm Cas}^{1/2}.
 \ee
 Eq.(\ref{cas-3}) has the solution
 \bea\label{rs-1}
 d^{1/2}_{\rm Cas}({a})=[C_0]^{-1/2} a ~~~\to~~~
 d_{\rm Cas}({a})=\frac{a^2}{C_0}.
 \eea
Comparing Eqs.~(\ref{1-3hh}), (\ref{rs-1}), one obtains
\be
C_0=H_0.
\ee
Thus, in our approach, the  parameter $C_0$ is equal to the
Hubble parameter $H_0$ which can be determined from observations.
Neglecting all matter effects, we obtain a new simple cosmological
model. Below we will show that the dominance of the rigid state
can persist even after an intensive creation of primordial particles.

\subsection{Hierarchy of cosmological scales
\label{hierarchy}}

In this Section we employ the Planck least action postulate to the
empty Universe action; define the initial value of the cosmological
scale factor, and consider  a hierarchy of cosmological scales in
correspondence with their conformal weights.

Let us consider the Early Universe  at the rigid state horizon
 (\ref{1-3hh}).
A hypothetical observer measures the conformal  horizon $d_{\rm hor}=2r_{\rm hor}(z)$
as  the distance that a photon covers within its light cone.
The latter is  determined by the zero interval equation
$d\eta^2-dr^2=0$ during
the photon lifetime in the homogeneous Universe, which is subject to the
condition $\eta_{\rm hor}=r_{\rm hor}(z)=1/[2H_0(1+z)^{2}]$, in accordance
with Eqs.~(\ref{ec-3}) and (\ref{rs-1}).
This means that
the four-dimensional space-time volume restricted by the horizon is equal to
 \be\label{Pl-1}
 V_{\rm hor}^{(4)}=\frac{4\pi}{3}r^3_{\rm hor}(z)\cdot \eta_{\rm hor}(z)=
 \frac{4\pi}{3\cdot16 H_0^4(1+z)^8}.
 \ee
It is natural to assume that at the instant of the Universe origin
the world was essentially quantum. In this case, the Universe action
can differ from the zero classical one by
the least action (or {\em quanta}), which presumably be small and be governed
by the Planck postulate of the least action  for  quantum
systems. Therefore, we suppose that action (\ref{1-3nt})
is subjected to the Planck's least action postulate at $a_{\rm Pl}=(1+z_{\rm  Pl})^{-1}$
 \bea
 \label{vac-a}
 W_{\rm Universe}=\rho_{\rm cr}V_{\rm hor}^{(4)}(a_{\rm Pl})=
 \frac{M_{\rm  Pl}^2}{H_0^2}\frac{1}{32 (1+z_{\rm  Pl})^{8}}=2\pi.
\eea
Using the present day $(\tau=\tau_0)$ observational data for the
Planck mass and  the Hubble parameter at $h\simeq 0.7$~\cite{Giovannini}
\bea\label{ri-12ah}
 M_{\rm C}\; e^{\langle D \rangle(\tau_0)}&=& M_{\rm Pl}=1.2211\cdot 10^{19} {\rm GeV},
 \qquad \langle D \rangle(\tau_0) = 0,
 \nonumber \\
\frac{d}{d\tau}\langle D \rangle(\tau_0)=H_0 &=& 2.1332\cdot 10^{-42} {\rm
GeV}\cdot h=1.4332 \cdot 10^{-42} {\rm GeV}\;,
 \eea
 we obtain from  (\ref{vac-a}) the primordial redshift value
 \bea\label{pl-2}
 a^{-1}_{\rm  Pl}=(1+z_{\rm  Pl}) \approx
 \left[{M_{\rm Pl}}/{H_0}\right]^{1/4}[4/\pi]^{1/8}/2\simeq 0.85\times 10^{15}.
 \eea
In other words, the Plank mass and the present day Hubble parameter value are
related to each other by the age of the Universe expressed in terms of the
cosmological scale factor.

In field theories characteristic scales, associated with physical states, are
classified according to the Poincar\'e group representation~\cite{logunov,wig}.
In our approach the Poincar\'e classification of energies arises from
the decomposition of the mean particle energy
$\omega_\tau =a^2\sqrt{{\bf k}^2+a^2M_0^2}$ conjugated to the dilaton time interval.
We express this decomposition in the form
 \be
 \label{ce-1}
 \langle\omega\rangle^{(n)}(a)=(a/a_{\rm  Pl})^{(n)}H_0\;,
 \ee
 based on Eq.~(\ref{pl-2},) where $\langle\omega\rangle^{(0)}_0=H_0$,
 $\langle\omega\rangle^{(2)}_0=k_0$,
$\langle\omega\rangle^{(3)}_0=M_0$,
$\langle\omega\rangle^{(4)}_0=M_{0\rm Pl}$.

  This equation enables one
 to introduce the conformal weights $n=0,2,3,4$ which  correspond to:
the dilaton velocity  $v_D=H_0$, the massless energy $a^2\sqrt{{\bf k}^2}$,
the massive one $M_0a^3$, and the Newtonian coupling constant $M_{\rm Pl}a^4$
(\ref{vac-a}), respectively.
One can also include in this classification   the scale of
 the nonrelativistic particle $H_0=a^{1}_{\rm  Pl}\times 10^{-13}$~cm$^{-1}$
 with the unit conformal weight of its energy $\omega^{\rm nonr.}_\tau=a^{1}{\bf
k}^2/M_0$.
As a result, the redshift leads to a hierarchy law of the present day ($a=1$)
cosmological scales
 \be
 \label{ce-2}
 \omega_0^{(n)}\equiv\langle\omega\rangle^{(n)}(a)\Big|_{(a=1)}=(1/a_{\rm
Pl})^{(n)}H_0
 \ee
shown in Table~\ref{Table:1}.

\begin{table}
\caption{The hierarchy law of the cosmological scales in GeV
         ($M^*_{\rm Pl}=\sqrt{3/(8\pi)}M_{\rm Pl}$).}
\label{Table:1}
\begin{tabular}{|c|c|c|c|c|c|}\hline
 n&{n}=0&{n}=1&{n}= 2&{n}=3&{n}=4
 \\ \hline
$\omega_0^{(n)}$ &$ H_0\!\simeq\! 1.4\cdot\!10^{-42}$&$R^{-1}_{\rm
Cel.S.}\!\simeq\!1.2\cdot \!10^{-27}\!$&$
  k_{0\rm CMB}\!\simeq\! 10^{-12}$&$\phi_{0}\!\simeq\!3\cdot 10^{2}$&$M^*_{\rm
Pl}\!\simeq\!4\cdot 10^{18}$\\ \hline
\end{tabular}
\end{table}

Table~\ref{Table:1} contains the scales corresponding to
the Celestial System size $(n=1)$,
the Cosmic Microwave Background mean wave-momentum $(n=2)$,
the electroweak scale of the SM $(n=3)$,
and the Planck mass $(n=4)$.
We  conclude that the observational data testify that the cosmic evolution
(\ref{ce-1})  of all these mean energies with conformal weights $(n=0,1,2,3,4)$
has a common origin which could be the Casimir vacuum energy.

{Thus, the application of the Planck least action postulate provides
the initial value $a_{\rm{Pl}}$ given by Eq. (\ref{pl-2}) in our model.
The Poincar\'e  classification of different states,
according to their conformal weights, reveals a
hierarchy of energy scales in agreement with observations.

\subsection{The exact solution of energy constraint in the CGR}

Let us consider the complete action~(\ref{1-3n}) in variables given by the Dirac-ADM
foliation. There are two treatments of the equation $N\delta W/\delta N=0$.

The first one belongs to Arnowitt, Deser and Misner~\cite{ADM}, who consider this equation
as the definition of the energy component of the total energy--momentum tensor
related to the Riemannian time $x^0$. This treatment leads to the concept of
non-localizable energy. However, the latter is not a diffeo-invariant quantity
and can not be associated with any observable, since
$x^0$ is the object of the diffeomorphisms~(\ref{1-1aa}).

 The second treatment belongs to Wheeler and DeWitt~(WDW)~\cite{WDW}, who consider this equation
as an algebraic first class energy constraint. Its resolution yields the  WDW  evolution generator.
This generator is identified with the canonical momentum of a time-like variable in the field space.

We conform to the rules of the second route. In particular, in our approach
the crucial step is the identification of this diffeo-invariant time-like evolution parameter in
the WDW  field space-time
with the zeroth  harmonic of the dilaton field~\cite{Barbashov:2005hu}.  The corresponding
canonical momentum is treated as the evolution generator (\ref{d-33})
in the Dirac-ADM Hamiltonian approach to the GR.
Recall that the zeroth and nonzeroth harmonics of the dilaton field are separated
by two projection operators:
the ``average'' $\langle D\rangle$ over the volume and the ``deviation'' $\overline{D}=D-\langle D\rangle$:
$D=\langle D\rangle+\overline{D}$ defined by Eqs.~(\ref{1-1dd})--(\ref{eta-m1}).
This projection removes the interference between
the independent degrees of freedom due to the orthogonality condition~(\ref{1-ortd});
for example, one has
 \be \nonumber
 \frac{1}{V_0}\int_{V_0} d^3x \biggl(\langle D\rangle + \overline{D}\biggr)^2 =
 \langle D\rangle^2 + \frac{1}{V_0}\int_{V_0} d^3x\; \overline{D}^2.
 \ee

Thus, the GR equations obtained by the variation of action~(\ref{1-3n}) after the separation
\bea\label{1-faa}
 && N \frac{\delta W_{\rm C}[D=\langle D\rangle+\overline{D}]}{\delta N}=
 \left\langle{N}^{-1}\left[{\partial_0 \langle D\rangle}\right]^2+
 {N}\left[{v_{\overline{D}} }\right]^2\right\rangle,
 \eea
differs from the equations obtained by the variation of this action before the separation
 \bea \label{1-fbb}
 && N \frac{\delta W_{\rm C}[D]}{\delta N}\Bigg|_{D=\langle D\rangle+\overline{D}}=\langle{N}
 \left[{v_D}\right]^2\rangle\Big|_{D=\langle D\rangle+\overline{D}}\,.
 \eea
Here $v_D$ is given by Eq.(\ref{1-3-d}) in Appendix A.
The logic of the second route requires that in the Hamiltonian approach we have to choose
the definite order of operations: the decomposition
and the variation of the action.

As a result,  the decomposition (\ref{1-1dd}) of the dilaton into two independent
harmonics (variables) requires the action to be a function of these two
independent variables.
The variation of the action with respect to the lapse function
$N\frac{\delta W_{\rm C}}{\delta{N}}=0$ gives
 \bea
 \label{h-c1d}
 \frac{[\partial_{\tau}{\langle D\rangle}]^2-\rho_\tau^{\rm v}}{\cal N}
  - {\cal N}\widetilde{{\cal H}}=0.
 \eea
Here we used Eqs.~(\ref{1-avd1f}) and (\ref{1-avd11}) to define
 \bea \nonumber
 && N\frac{\delta }{\delta{N}}\;\frac{1}{N_0}=  N\frac{\delta }{\delta{N}}
 \left[\frac{1}{V_0} \int_{V_0}d^3x \frac{1}{N}\right]
 = -\frac{1}{N} = -\frac{1}{{\cal N}\; N_0},
 \\ \nonumber
 && N\frac{\delta }{\delta{N}}\; N_0=  \frac{N_0^2}{N} = \frac{N_0}{\cal N},
 \qquad \frac{[d\langle D\rangle]^2}{N_0^2\; (dx^0)^2} = [\partial_{\tau}{\langle D\rangle}]^2,
 \eea
where we dropped arguments for simplicity.
Eq.~(\ref{h-c1d}) has the additional term
$([\partial_{\tau}{\langle D\rangle}]^2-\rho_\tau^{\rm v})/{\cal N}$
in the comparison with the original Einstein equation ${\cal N}\widetilde{{\cal H}}=0$
~\cite{lif,Mukhanov:1990me}.

Constraint (\ref{h-c1d}) determines the diffeo-invariant lapse function
 \be
 \label{CC-2}
 {\cal N}=\frac{\left\langle \sqrt{\widetilde{{\cal H}}}\right\rangle}
 {\sqrt{\widetilde{{\cal H}}}}
 \ee
by means of the Hamiltonian density
 $ \widetilde{{\cal H}}=-\frac{2}{3}e^{-7D/2}\triangle e^{-D/2}+{\cal H}_g$
(see Eq. (\ref{CC-3}), (\ref{3g})) and its spatial average
$\langle \sqrt{\widetilde{{\cal H}}}\rangle
=V^{-1}_0\int d^3x \sqrt{\widetilde{{\cal H}}}$.
The additional term solves problems of the GR associated with the unambiguous
definition of the energy and the lapse function~\cite{Barbashov:2005hu}.
Moreover, it leads also to novel physical consequences for the large-scale structure
of the Universe,
discussed in detail in the next Section and in Appendix B.

In virtue of this result, by averaging over spatial volume Eq.(\ref{h-c1d}),
we obtain the equation for the WDW evolution generator~(\ref{d-33}):
 \bea\label{CC-1}
 (\partial_{\tau}\langle D\rangle)^2 \equiv \frac{1}{4} P_{\langle D\rangle}^2
 =\rho_\tau^{\rm v}+{\textsf{H}}_\tau^{g}/V_0
 =\rho^g_{\rm tot}\;.
 \eea
Here $ \textsf{H}_\tau^g$ is the Hamiltonian constructed with the aid of two last
terms of action~(\ref{1-3n}).

The solution of Eq.~(\ref{CC-1}) provides the exact time-redshift relation
 \bea\label{Dirac-c2}
 \tau &=&\int\limits_{\langle D \rangle_I}^{\langle D \rangle_0}
 d\langle D \rangle \left[\rho^g_{\rm tot}\right]^{-1/2}.
 \eea

Thus, the Hamiltonian approach to the  CGR provides the exact solution of
the energy constraint in terms of the conformal
field variables connected with the Einstein ones by the scale transformation
 \be\label{sc-1}
 \widetilde{F}^{(n)}=e^{nD}F^{(n)},
 \ee
where $(n)$ conformal weights $(n=-1,0,3/2,2)$
for scalars, vectors, spinors, and tensors, respectively.

The explicit solution of the constraint,
Eq.~(\ref{CC-2}), results in the constraint-shell interval
 \bea\label{ds-1}
 \widetilde{ds}^2\!\!&=&\!\! e^{-4D}\!\frac{\langle\sqrt{{\cal H}}\rangle^2}
 {\cal H}\!d\tau^2-\left(
 \widetilde{{\bf e }}_{i(a)}dx^i -
 \overline{\cal N}_{(a)} d\tau\!\right)^2\!\!.
 \eea
>From the requirement that the squared time interval is a positive definite
it follows that we deal with a field theory with positive-definite metrics of
fields in the Hilbert space ${\cal H}>0$. This positive-definite metrics
is emerged due to condition (\ref{D-2a}) which is a result of the dilaton decomposition
Eq.~(\ref{1-1dd}).

The basic cosmological problems are to solve the Hamiltonian equations of motion
with respect to the dilaton
${\langle D \rangle}$ and to establish the relation (\ref{Dirac-c2}) with
observational quantities.

We stress that the solution~(\ref{Dirac-c2}) of the energy constraint~(\ref{h-c1d}) is
analogous to the corresponding Einstein equation obtained in the homogeneous approximation.
Note that Eq.~(\ref{CC-2}) defines the relation between the lapse function
and matter, see Appendix~B and Ref.~\cite{Arbuzov:2010fz}.

\section{Affine gravitons and their properties
\label{sect_CC}}

\subsection{Affine graviton}

Let us consider the graviton action~(\ref{1-3ng}) in order to resolve
the constraints arising due to invariance of the action under the general
coordinate transformations ~(\ref{1-1b}) (\textit{i.e.} diffeo-invariance).

In the constraint-shell interval~(\ref{ds-1}) only the simplex components
$\overline{\omega}_{(a)}(d)=\widetilde{{\bf e }}_{i(a)}dx^i$ are constrained variables.
They obey the condition of the
diffeo-invariance. It is one of the main differences of the CGR from the GR.

The choice of the symmetry condition in the CGR leads to the result that follows from the
theorem~\cite{Tod}: \textit{any arbitrary two-dimensional space metric
$dl^2=h_{AB}dx^A dx^B$,\ $(A,B=1,2)$, can be represented by diffeomorphisms
$x^A\to \widetilde{x}^A=\widetilde{x}^A(x^1,x^2)$ in a diagonal form.} The
result consists in the fact that a kinemetric-invariant nonlinear plane wave
moving in the direction ${\bf k}$ with the unit determinant $det\;h\;=1$
contains only a single metric component.

In the  frame of reference  ${\bf k}=(0,0, k_3)$
one has $\widetilde{{\bf e}}^1_{(1)}=e^{g(x_{(3)},\tau)}$,
$\widetilde{{\bf e}}^2_{(2)}=e^{g(x_{(3)},\tau)}$,
and $\widetilde{{\bf e}}^3_{(3)}=1$;
all other (non-diagonal) components $\widetilde{{\bf e}}^i_{(a)}$ are equal to zero.
Thus, we obtain
 \bea \label{dg-2a2}
 \overline{{\omega}}_{(1)}&=& 
 dX_{(1)}- [X_{(1)}]d g,
 \\\label{dg-2a3}
 \overline{{\omega}}_{(2)}&=&
 dX_{(2)} + [X_{(2)}]d g,
 \\
 \label{dg-2ax}
 \overline{{\omega}}_{(3)}&=&dx_{3}=dX_{(3)},
 \eea
where a single-component affine graviton $g=g(X_{(3)},\tau)$  is a function depending
on the time and a single spatial coordinate $X_{(3)}$
in the tangent space $X_{(b)}$. The solutions of the
equation $\frac{\delta W}{\delta g}=0\to g=g(\eta, X)$ can be expressed via the
tangent coordinates:
   \bea
   \label{X-1}X_{(1)}= e^{g(x_{(3)},\tau)}  \overline{x}^1
   \\
   \label{X-2} X_{(2)}= e^{-g(x_{(3)},\tau)}  \overline{x}^2. \eea

Eqs.~(\ref{dg-2a2}) and (\ref{dg-2a3}) mean
 an expansion (or contraction) of the hypersurface $X_{(A)}\,( A=1,2)$ perpendicular
to the direction of the  gravitational wave propagation $X_{(3)}$.
A gravitation wave changes the particle velocity   via the Hubble like law:
the more base, the more  additional velocity induced by the graviton.

The exact local Hamiltonian density for the affine graviton is given by (\ref{3g})
\bea\label{3g-1}
 {\cal H}_{\rm g}&=&\left[6{P^2_{(a)(b)}}+
 \frac{1}{6}R^{(3)}(\widetilde{{\bf e}})\right]\;,
 \eea
where $R^{(3)}({{\bf e}})$ and $P^2_{(a)(b)}$  are defined by Eqs.~(\ref{1-17}) and (\ref{1-17g}),
respectively. For the frame of reference ${\bf k}=(0,0, k_3)$,
 we have~\cite{ll}:
 \bea \label{dg-11}
 R^{(3)}(\widetilde{{\bf e}})&=&(\partial_{(3)}g)^2,\qquad
 {P^2_{(a)(b)}}=\frac{1}{9}\left[\partial_\tau{g}
 \right]^2.
 \eea
There is a difference of the diffeo-invariant affine graviton from the
a metric graviton $g^{TT}_{ij}=g^{TT}_{ji}$ in the GR~\cite{ll}.
While the affine graviton has  a single degree of freedom,
the metric graviton has two  traceless  and
transverse components that satisfy four constraints
 \bea\label{TT-1}
 g^{TT}_{ii}&=&0,\\\label{TT-2}
 g^{TT}_{i3}&=&g^{TT}_{3i}=0.
 \eea

In general case of the CGR $\widetilde{{\bf e}}_{(b)i}={\bf e}^{T}_{(b)i}$,
both the transverse constraint
 \be\label{tr-1}
 \partial_i{\bf e}^{T}_{(b)i}=0
 \ee
and the unit determinant one
 \be\label{tr-2}
 |{\bf e}^{T}_{(b)i}|=1
 \ee
(as the analog of the Lichnerowicz gauge in the metric formalism~\cite{lich})
admit to generalize  Eqs. (\ref{dg-2a2}), (\ref{dg-2a3}), and (\ref{dg-2ax}) for the
linear forms,
 \bea
 \label{P-1}
 {\overline{\omega}}_{(b)}(d)&=&{\bf e}^{T}_{(b)i}dx^i
 \\\nonumber
 &=&d [{\bf e}^{T}_{(b)i}x^i]-x^jd{\bf e}^{T}_{(b)j}\\\nonumber
 &=& dX_{(b)}-X_{(c)}{\bf e}^{Ti}_{c}d{\bf e}^{T}_{(b)i}\\\nonumber
 &=&dX_{(b)}-X_{(c)}\left[\omega^R_{(b)(c)}+
 \omega^L_{(b)(c)}\right]
 \eea
in the tangent coordinate space.  Here $X_{(b)}$  can be obtained by the formal
generalization
of Eqs. (\ref{dg-2a2}), (\ref{dg-2a3}), and (\ref{dg-2ax})
by means of the Leibniz rule
${\bf e}^{T}_{(b)\,i}d[x^i]=
d[{\bf e}^{T}_{(b)\,i}x^i] -x^id{\bf e}^{T}_{(b)i}$.
 The diffeomorphism-invariance
admits the choice of the gauge in Eq. (\ref{P-1})
 \be\label{gauge-1}
 \omega_{(b)\,(c)}^L=0.
 \ee

Similar result is valid for a general case of arbitrary wave vector
$\textbf{k}=\frac{2\pi}{V^{1/3}_0}\textbf{l}$, where
$X_{(3)}$ is replaced by $X_{(k)}=\textbf{k}\textbf{X}/\sqrt{{\textbf{k}}^2}$.
The single-component graviton  $g(\tau,\textbf{X})$ considered
as the tensor massless representation of the Wigner classification
of the Poincar\'e group~\cite{logunov} can be decomposed
into a series of strong waves (in natural units)
 \bea \label{R-1a}
 \omega_{(a)(b)}^R(\partial_{(c)})
 =i\sum_{\textbf{k}^2\not=0}
 \frac{e^{i\textbf{\textbf{k}\textbf{X}}}}{\sqrt{2\omega_{\textbf{k}}}}
 \textbf{k}_{c}[\varepsilon^R_{(a)(b)}(k)
 g_{\textbf{k}}^+(\eta)+\varepsilon^R_{(a)(b)}(-k)
 g_{-\textbf{k}}^-(\eta)]\;.
 \eea
 Here ${\varepsilon}^R_{(a)\,(b)}(\textbf{k})$  satisfies the constraints
 \bea\label{TT-3}
 \varepsilon^R_{(a)(a)}(k)&=&0,\\\label{TT-4}
 {\bf k}_{(a)}\varepsilon^R_{(a)(b)}(k)&=&0\;,
 \eea
similar to (\ref{TT-1}), (\ref{TT-2}). The variable
$\omega_{\textbf{k}}=\sqrt{{\textbf{k}}^2}$ is the graviton energy and the affine
graviton
 \be \label{3-1pp}
 \overline{g}_{\textbf{k}}=\frac{\sqrt{8\pi}}{M_{\rm Planck}V^{1/2}_0}g_{\textbf{k}}
 \ee
is normalized  to the units of volume and time (like a photon in QED~\cite{ll}).

In the mean field approximation
 \bea \label{mfa-11}
 {\cal N}(x^0,x^j)&=& 1, ~~~N^j=0, ~~~~\overline{D}=0,\\
 \label{mfa-6a}
 \widetilde{ds}^2&=&[d\eta]^2-[\omega_{(b)}\otimes\omega_{(b)}],
 \eea
 when one neglects all Newtonian--type
interactions, the action of an affine graviton
reduces to the form of the exact action for the strong gravitational wave~\cite{ll}
 \bea
 \label{2-3ngw}
 {W}^g_{\rm lin}&=&\int\limits_{}^{}d\tau\textsf{L}^g_{\tau}, \\
 \label{2-3ngg2}
  \textsf{L}^g_{\tau}&=&\frac{v^2_{(a)(b)}
 -\,e^{-4D}\!R^{(3)}}{6}
 = \sum\limits_{\textbf{k}^2\not =0}\frac{
 v^g_{\textbf{k}}v^g_{-\textbf{k}}-\,e^{-4D} \textbf{k}^2
 \overline{g}_{\textbf{k}}\overline{g}_{-\textbf{k}}}{2}
\nonumber \\
 &=&\left[\sum\limits_{\textbf{k}^2\not =0}
 p_{-\textbf{k}}^gv^g_{\textbf{k}}\right]-\textsf{H}^g_{\tau},
 \eea
where $v^g_{\textbf{k}}=\partial_\tau \overline{g}_{\textbf{k}}$ is the derivative
with respect to the dilatonic time interval (\ref{eta-1}) and
 \bea\label{2-10gh}
 \textsf{H}^g_{\tau}=\sum\limits_{\textbf{k}^2\not =0}
 \frac{p_{\textbf{k}}^gp_{-\textbf{k}}^g+ e^{-4\langle D\rangle}
 \textbf{k}^2 \overline{g}_{\textbf{k}}\overline{g}_{-\textbf{k}}}{2}
 \eea
 is the corresponding Hamiltonian.

Thus, in the mean field approximation~(\ref{mfa-11}) the diffeo-invariant sector
of the strong gravitational plane waves coincides with a bilinear theory given
by Eqs.~(\ref{2-3ngw}) -- (\ref{2-10gh}). In
this approximation our model is reduced to a rather simple theory which is
bilinear with respect to the single-component graviton field
as discussed also in Ref.~\cite{Arbuzov:2010fz}.
Note that we consider here the tangential space, and the chosen variables allow
to obtain the simple solutions. The main  postulated condition here was
the requirement of the diffeo-invariance of the graviton equation of motion.
While in the standard GR the symmetry properties are required only for the interval,
we impose  the symmetry with respect to diffeomorphisms also on the Maurer--Cartan forms.

\subsection{Comparison with metric gravitons}

It is instructive to compare the properties  of the affine and metric
gravitons, which was done first in Ref.~\cite{Arbuzov:2010fz}).

The action of metric gravitons in the accepted GR~\cite{Gr-74,Babak:1999dc}
coincides with the affine one (\ref{2-3ngw}) in the lowest order of the
decomposition over  ${\bf k}^2/M^2_{\rm Pl}$
 \bea
 \label{2-3nh}
 W^{\rm GR}_{\rm non-lin}&=&{W}^g_{\rm lin}+{W}_{\rm non-lin},
 \eea
if we keep
only diagonal graviton components. It is well-known~\cite{Faddeev:1973zb}
that the accepted action (\ref{2-3nh})  is
highly nonlinear even in the approximation~(\ref{mfa-11}).

In the approximation  (\ref{mfa-11}),
we keep
only the dynamical part $\omega^R_{(cb)}$ (which enters into the action~(\ref{2-3ngw}))
and the present day value of the cosmological scale factor $a=e^{-\langle D\rangle}=1$.
Let us compare the affine gravitons (\ref{P-1})
 with the commonly accepted  metric gravitons, given by the decomposition
\cite{Gr-74,Babak:1999dc}
 \bea
 \label{fs-1s}
 \widetilde{ds}_h^2&=&(d\eta)^2  - dx^idx^j
\left(\delta_{ij}+2h^{TT}_{ij}+\ldots \right).
 \eea
In the accepted case, the graviton moves in the direction of vector
$\textbf{k}$, its wave amplitude $\cos\{\omega_{\textbf{k}}x_{(k)}\}$ depends
on the scalar product $x_{(k)}=(\textbf{k}\cdot
\textbf{x})/\omega_{\textbf{k}}$.

The graviton changes the squared test
particle velocity
$\left(\frac{ds}{d\eta}\right)^2\sim \frac{dx^idx^j}{d\eta
\,d\eta}\varepsilon^\alpha_{ij}$ in the plane, orthogonal to the direction of
motion. Here $\varepsilon^\alpha_{ij}$ is the traceless transverse tensor:
$\varepsilon^\alpha_{ii}=0$ and $k_i\varepsilon^\alpha_{ij}=0$.
All these effects are produced by the first order of series (\ref{fs-1s})
 \bea
 \label{pl-1s}
 dl_{h}^2 &=& 2dx^idx^j h^{TT}_{ij}(t,\textbf{x})
 \nonumber \\
 &=& dx^idx^j\varepsilon^\alpha_{ij}
 \sqrt{6}\cos\{\omega_{\textbf{k}}x_{(k)}\}({H_0}/{\omega_{\textbf{k}}})
 {\Omega_{\textbf{k}\rm h}^{1/2}}+O(h^2),
 \eea
where  $H_0$ is the Hubble parameter, $\Omega_{\textbf{k}\rm
h}=\omega_{\textbf{k}}N_{\textbf{k}\rm h}/[ V_0\rho_{\rm cr}]$ is the energy
density of the gravitons in units of the cosmological critical energy density (\ref{cr-1a}).
One observes that in the accepted perturbation theory the  contribution of a
single gravitational wave to the geometrical intervals, Eq.(\ref{pl-1s}), is
suppressed by the factor $H_0/\omega_{\textbf{k}}$.

In our version the linear term of the spacial part of Eq.~(\ref{P-1})
takes
the form
 \bea \label{2-1gg1} \nonumber
 dl_{g}^{2} = 2dX_{(b)}X_{(c)} \omega^R_{(c)(b)}
 = dX_{(b)}X_{(c)}
 \varepsilon^\alpha_{(c)(b)}
 \sqrt{6}\cos\{\omega_{\textbf{k}}X_{(k)}\}{H_0} {\Omega_{\textbf{k}\rm h}^{1/2}}.
 \eea
Evidently, two models (the GR and the CGR) differ by
an additional factor which can be
deduced from the ratio
 \be\label{ratio}
 \bigg|\frac{dl_h^2 }{dl_g^2}\bigg| =
 \bigg|\frac{ dx^idx^j\left(h^{TT}_{ij}\right)}
 {(dX_{(b)}X_{(c)}\omega^R_{(c)(b)})}\bigg|
 \simeq \frac{1}{r_{\!\!\bot}\,\omega_{\textbf{k}}}
\sim \frac{\lambda_g}{r_{\!\!\bot}}.
 \ee
Here $r_{\!\!\bot}\,=\sqrt{|\vec X_{\!\!\bot}\,|^2}$ is the coordinate
distance between two test particles  in the plane perpendicular to the wave
motion direction and $\lambda_g$ is the graviton wave length. Therefore, in the CGR
there is the effect of the  \textit{expansion} of the plane perpendicular to the affine wave
motion direction.

As a result, in the CGR the total velocity of a test
classical particle in the central gravitational field of a mass $M$ and of a strong
gravitational wave is the sum of three velocities
at the cosmic evolution $a\neq 1$. The first term is the
standard Newtonian (N) velocity, the second is the velocity of the {\em graviton
expansion} (g) in the field of a gravitational wave, and the third one is the
velocity of the Hubble  evolution (H):
 \be \label{sum-1}
|\vec v|^2=\big|\frac{{dl}_g}{d\eta}\big|^2=
\left[\underbrace{{\vec n}_{\rm N}\sqrt{\frac{r_g}{2R_{\bot}}}}_{\rm
Newtonian~ velocity} + \underbrace{{\vec n}_{\rm g}
\sqrt{R_{\bot}H_0}\sqrt{\Omega_g}}_{\rm graviton~ expansion} +
\underbrace{{\vec n}_{\rm H}\gamma H_0R_{\bot} }_{\rm Hubble~ evolution}
\right]^2\!\!\! .
\ee
Here, $R_{\bot}=r_{\bot}a(\eta)$ is the Friedman distance from the central
mass, $H_0$ is the Hubble parameter, $r_g(R_{\bot})=M/M_{\rm Pl}^2$ is a constant
gravitational radius,
and
 \be \label{vectors1} \left\{
 \begin{array} {l}
 {\vec n}_{\rm N}=(0,-1,0),\\
 {\vec n}_{\rm g}=(+1/\sqrt{2},-1/\sqrt{2},0),\\
 {\vec n}_{\rm H}=(1,0,0)\\
 \end{array}\right.
 \ee
are the unit velocity vectors. Their scalar products are ${\vec n}_{\rm
N}\cdot {\vec n}_{\rm g}\not=0$, ${\vec n}_{\rm N}\cdot {\vec n}_{\rm H}=0$,
${\vec n}_{\rm N}\cdot {\vec n}_{\rm g}\not=0$, and ${\vec n}_{\rm N}\cdot
{\vec n}_{\rm H}=0$. The graviton energy density $\Omega_g$ is given in units
of the cosmological critical energy density  $\rho_{\rm cr.}$

The last two terms provide possible sources of a modified Newtonian dynamics.
One observes that the interference of the Newtonian and the graviton-induced
velocities in (\ref{sum-1}) $v_{\rm n-g~interf}\simeq \sqrt[4]{\Omega_g r_g H_0}$
does not depend on the radius $R_{\bot}$.

The third term could imitate the Dark Matter effect in  COMA-type clusters
with $|R|\sim 10^{25}$cm, in accordance with the validity limit of the
Newtonian dynamics, $r_g/R_{\rm limit}<2(R_{\rm limit}H_0)^2$, discussed
in~\cite{Einstein:1945id,Gusev:2003sv}. The factor $\gamma=\sqrt{2}$ is defined by the
cosmological density~\cite{z-06}.

Thus, in our model strong gravitational waves  possess peculiar properties
which can be tested by observations and experiments.

\subsection{Vacuum creation of affine gravitons}

Here we are going to study the effect of intensive creation of affine gravitons.
We will briefly recapitulate the derivation given in Ref.~\cite{Arbuzov:2010fz} and further,
using the new results of Sect.~\ref{hierarchy}, estimate the number of created
particles.

The approximation defined by Eqs.~(\ref{2-3ngw})--(\ref{2-10gh})
can be rewritten by means of the conformal variables and coordinates,
where the action
\bea\label{W-1c}
 {W}^g_{\rm lin}=\int\limits^{\eta_0}_{\eta_I} d\eta \left[-V_0{(\partial_\eta
 \langle D\rangle)^2} e^{-2\langle D\rangle}+\textsf{L}^g_{\rm \eta}\right]
 \eea
is given in the interval $\eta_I \leq \eta \leq \eta_0$ and spatial
volume $V_0$. Here the Lagrangian and Hamiltonian
 \bea \label{2-3ng}
 \textsf{L}^g_{\rm \eta}&=& \sum\limits_{\textbf{k}^2\not =0}\,e^{-2 \langle
 D\rangle}\frac{ v^g_{\textbf{k}}v^g_{-\textbf{k}}- \textbf{k}^2
 \overline{g}_{\textbf{k}}\overline{g}_{-\textbf{k}}}{2}
 =\left[\sum\limits_{\textbf{k}^2\not =0}
 p_{-\textbf{k}}^gv^g_{\textbf{k}}\right]- \textsf{H}^g_{\rm \eta},
 \\ \label{2-3ngh1}
 \textsf{H}^g_{\rm \eta}&=&\sum\limits_{\textbf{k}^2\not =0}\,
 \frac{ e^{2 \langle D\rangle}p^g_{\textbf{k}}p^g_{-\textbf{k}}
 + e^{-2 \langle D\rangle}{\omega^2_{0\textbf{k}}}
 \overline{g}_{\textbf{k}}\overline{g}_{-\textbf{k}}}{2}
 \eea
are defined in terms of the variables $\overline{g}_{\textbf{k}}$,
their momenta, and one-particle conformal energy
 \be \label{hh-1}
 p_{\textbf{k}}^g=e^{-2\langle D\rangle}v^g_{\textbf{k}}
 = e^{-2 \langle D\rangle}\partial_\eta
 \overline{g}_{\textbf{k}},\quad
 {\omega}^g_{0{\bf k}}=\sqrt{{\bf k}^2},
 \ee
 respectively. The transformation (squeezing)
 \be \label{2-10gh1}
 p_{\textbf{k}}^g=\widetilde{p}_{\textbf{k}}^g e^{-\langle D\rangle}
 [\omega^g_{0{\bf k}}]^{-1/2}, \qquad \overline{g}_{\textbf{k}}
 = \widetilde{g}_{\textbf{k}} e^{\langle D\rangle}[\omega^g_{0{\bf k}}]^{1/2}
 \ee
leads to the canonical form
 \bea \label{2-newh}
 \textsf{H}^g_{\rm \eta}&=&\sum\limits_{\textbf{k}^2\not =0}
 \omega^g_{0{\bf k}}\frac{\widetilde{p}^g_{\textbf{k}}\widetilde{p}^g_{-\textbf{k}}
 \!+\!
 {\widetilde{g}}_{\textbf{k}}{\widetilde{g}}_{-\textbf{k}}}{2}=
 \sum\limits_{\textbf{k}}^{}\underline{{\cal H}}_{\textbf{k}}^g,
 \\ \label{3-8ac}
 \underline{{\cal H}}_{\textbf{k}}^g&=&
 \frac{\omega^g_{0\textbf{k}}}{2}
 [\widetilde{g}_{\textbf{k}}^+\widetilde{g}_{-\textbf{k}}^-\!+\!
 \!\widetilde{g}_{\textbf{k}}^-\widetilde{g}_{-\textbf{k}}^+]\;,
 \eea
where
 \be \label{hv-1}
 \widetilde{g}^{\pm}_{\textbf{k}}=\left[\widetilde{g}_{\textbf{k}}
 \mp i\widetilde{p}_{\textbf{k}} \right]/\sqrt{2}
 \ee
are the conformal-invariant
classical variables in the holomorphic representation~\cite{z-06,Pervushin:1999mq}.

In virtue of Eqs.~(\ref{hh-1})--(\ref{hv-1}), the action (\ref{W-1c}) takes
the form
 \bea\label{W-3c}
 {W}^g_{\rm lin} &=& 
 \int\limits^{\eta_0}_{\eta_I} d\eta \left[ -{V_0(\partial_\eta \langle
 D\rangle)^2} e^{-2\langle D\rangle}
 - \textsf{H}^g_{\rm \eta}\right]
 \nonumber \\
 &+&\int\limits^{\eta_0}_{\eta_I}d\eta
 \sum\limits_{\textbf{k}^2\not =0} \widetilde{p}_{-\textbf{k}}
 \left[\partial_\eta \widetilde{g}_{\textbf{k}}+\partial_\eta\langle D\rangle
 \widetilde{g}_{\textbf{k}} \right].
 \eea
The evolution equations for this action are
 \be\label{cc-5}
 \partial_\eta \widetilde{g}^{\pm}_{\textbf{k}}=\pm i \omega^{g}_{0\textbf{k}}
 \widetilde{g}^{\pm}_{\textbf{k}}+H_\eta\,\widetilde{g}^{\mp}_{\textbf{k}},
 \ee
where
 $H_\eta=\partial_\eta(\ln a)=-\partial_\eta\langle D\rangle$
is the conformal Hubble parameter (in our model $H_\eta=H_0/a^2$).

It is generally accepted to solve these equations by means of the Bogoliubov
transformations
 \bea\label{cc-6}
 \widetilde{g}^{+}_{\textbf{k}}&=&\alpha_{\textbf{k}} b_{\textbf{k}}^++
 \beta^*_{\textbf{-k}} b_{\textbf{-k}}^-,
 \\\label{cc-6c}
 \widetilde{g}^{-}_{\textbf{k}}&=&\alpha^*_{\textbf{k}} b_{\textbf{k}}^-+
 \beta_{\textbf{-k}} b_{\textbf{-k}}^+,
 \\\label{cc-6cc}
 \alpha_{\textbf{k}}&=&\cosh r^g_{\textbf{k}}e^{i\theta^g_{\textbf{k}}},~~~~~
 \beta^*_{\textbf{k}}=\sinh r^g_{\textbf{k}}e^{i\theta^g_{\textbf{k}}},
 \eea
where $r^g_{\textbf{k}}$ and $\theta^g_{\textbf{k}}$ are the squeezing
parameter and the rotation one, respectively (see for details reviews~\cite{Gr-74,grib80}).
These transformations preserve the Heisenberg algebra
$O(2|1)$~\cite{jor-63} and diagonalize Eqs.~(\ref{cc-5}):
 \bea\label{cc-7}
 \partial_{\eta}b^{\pm}_{\textbf{k}}=
 \pm i \overline{\omega}^g_{B\textbf{k}} b^{\pm}_{\textbf{k}},
 \eea
if the parameters of squeezing $r^g_{\textbf{k}}$ and rotation
$\theta^g_{\textbf{k}}$ satisfy the following equations~\cite{z-06}:
 \bea \label{cc-8}
  \partial_\eta{r^g}_{\textbf{k}}&=&H_\eta\cos 2\theta^g_{\textbf{k}},
 \\ \label{cc-9}
 {\omega}^g_{0\textbf{k}}-\partial_\eta{\theta^g}_{\textbf{k}}
 &=& H_\eta\coth 2r^g_{\textbf{k}}\sin 2\theta^g_{\textbf{k}},\\
 \label{cc-9b}
 {\omega}^g_{B\textbf{k}}&=&\frac{{\omega}^g_{0\textbf{k}}-
 \partial_\eta{\theta^g}_{\textbf{k}}}{\coth 2r^g_{\textbf{k}}}.
 \eea
A general solution of the classical equations can be written with the aid of a
complete set of the initial data $ b^{\pm}_{0\textbf{k}}$:
 \bea \label{cc-10g}
 b^{\pm}_{\textbf{k}}(\eta)=\exp\left\{\pm i \int\limits^{\eta}_{\eta_0}
 d\overline{\eta}{\,\omega}^g_{B\textbf{k}}(\overline{\eta})\right\}
 b^{\pm}_{0\textbf{k}}.
 \eea

On the other hand, quantities $ b^{+}_{0\textbf{k}}( b^{-}_{0\textbf{k}})$ can
be considered as the creation (annihilation) operators, which satisfy the
commutation relations:
 \be
 [b^{-}_{0\textbf{k}},b^{+}_{0\textbf{k}'}]=\delta_{\textbf{k},\textbf{-k}'},\quad
 [b^{-}_{0\textbf{k}},b_{0\textbf{k}'}^{-}]=0,\quad
 [b^{+}_{0\textbf{k}},b^{+}_{0\textbf{k}'}]=0, \label{bcom}
 \ee
if one
introduces the   vacuum state as $b^{-}_{0\textbf{k}}|0\rangle=0$. Indeed,
relations (\ref{bcom}) are the results of: i) the classical Poisson
bracket $\{P_{\widetilde{F}},\widetilde{F}\}=1$
which transforms into
 \be
 [\widetilde{g}^{-}_{\textbf{k}},\widetilde{g}^{+}_{-\textbf{k}}]
 =\delta_{\textbf{k},\textbf{k}'};
 \label{pcom}
 \ee
ii) the solution (\ref{cc-10g}) for the initial data;
iii) the Bogoliubov transformations (\ref{cc-6}), (\ref{cc-6c}).

With the aid of Eqs.~(\ref{cc-6})-(\ref{cc-6cc}) and
(\ref{cc-10g})-(\ref{pcom}) we are able to calculate the  vacuum expectation
value of the total energy (\ref{2-newh}),(\ref{3-8ac})
 \be
 \label{H-2ad}
 \langle 0|\textsf{H}_\eta^g(a)|0\rangle=
\sum\limits_{{\bf k}}^{}{\omega}^g_{0\textbf{k}}|\beta_{\textbf{k}}|^2=
 \sum\limits_{{\bf k}}^{}
 {\omega}^g_{0\textbf{k}}
 \frac{\cosh\{2r_{\bf k}^g(a)\}-1}{2}.
 \ee

The numerical analysis~\cite{Arbuzov:2010fz}
 of Eqs.~(\ref{cc-8})-(\ref{cc-9}) for unknown variables $(r^g_{\textbf{k}},
\theta^g_{\textbf{k}})$
 with the zero boundary conditions at $a=a_{I}$ (at the beginning of creation)
 \bea \label{3-9bb0h}
  {r^g}_{\textbf{k}}(a_I)=0,
\qquad \theta^g_{\textbf{k}}(a_I)=0
 \eea
  enables us to suggest an
approximate analytical solution for the evolution equations.

Our approximation consists in the following. It arises, if instead of
$r_{\textbf{k}}$ one substitutes an approximate value $r_{\rm apr}$ in the
vicinity of  the soft mode of the  Bogoliubov energy (\ref{cc-9b})
$\omega_{0\rm appr}=\partial_\eta{\theta^g}_{\rm appr}$,
 \bea \label{c-4cn1}
 r_{\rm appr}&=&\frac{1}{2}
 \int\limits_{X_I=2\theta^g_{\rm appr}(a_I)}^{X=2\theta^g_{\rm appr}(a)}
 \frac{d\overline{X}}{\overline{X}}
 {\cosh \overline{X}}\simeq 2\langle D\rangle_I,
 \\ \label{c-4cn12}
 X(a)&=&2\theta^g_{\rm appr}(a)=
 2\int\limits_{\eta(a_I)}^{\eta(a)}d\eta \omega_{0\bf k}.
 \eea
 This  soft
mode  provides a  transition~\cite{Arbuzov:2010fz}  at the point $a^2_{\rm relax}\simeq
2a^2_{\rm Pl}$ from the unstable state of the particle creation to the stable
state with almost a constant occupation number  during the relaxation time
 \be\label{relax-1}
 \eta_{\rm relax}\simeq 2e^{-2\langle
 D\rangle_I}/(2H_0)\equiv 2a_I^2/(2H_0).
 \ee
 At the point of the relaxation,
the determinant of Eqs.~(\ref{cc-5}) changes its sign and becomes
positive~\cite{Andreev:1996qs}. Finally, we obtain
 \bea
 \label{3-8v}
  \langle 0|{\cal H}_{\textbf{k}}^g|0\rangle\big|_{(a>a_{\rm relax})} \!=\!
\omega^g_{0\textbf{k}}\frac{\cosh[2r_{\textbf{k}}^g]-1}{2}
  \approx \frac{\omega^g_{0\textbf{k}}}{4a^4_I}.
 \eea
 We have verified that the deviation of the results obtained with the aid of
 this formula  from  the numerical
solutions of  Eqs.~(\ref{cc-8})--(\ref{cc-9}) (see Ref.~\cite{Arbuzov:2010fz}) does
not exceed 7\%.

In virtue of this result, we obtain the total energy
 \bea\label{H-2a} \!\!
 \langle 0|\textsf{H}_\eta^g|0\rangle\big|_{(a>a_{\rm relax})}\!\! \approx
 \frac{1}{2a_I^4}\sum\limits_{{\bf k}}^{}\frac{{\omega}^g_{0\textbf{k}}}{2}
 \equiv \frac{\textsf{H}^g_{\eta\; {\rm Cas}}({a})}{2a_I^4},
 \eea
 where $\textsf{H}^g_{\eta\; {\rm Cas}}({a})$
 is the  Casimir vacuum energy (\ref{cr-1a})~\cite{grib80,sh-79}.

Thus, the total energy of the created gravitons is
 \bea\label{H-2ab}
 \langle 0|\textsf{H}_\eta^g|0\rangle \simeq
 \frac{{\widetilde{\gamma}}H_0}{4a^2a^4_I}.
 \eea
 It appeared that the dilaton initial data $a_I=e^{-\langle D\rangle_I}$
and $H_0$ determine both the total energy  (\ref{H-2a}) of the created gravitons
and their occupation number $N_g$ at the relaxation time (\ref{relax-1}):
\bea\label{H-2abr}
 N_g(a_{\rm relax})\simeq \frac{\langle
0|\textsf{H}_\eta^g|0\rangle}{\langle\omega^g_k\rangle} \simeq
 \frac{\widetilde{\gamma}^{(g)}}{16 a^6_I}\simeq 10^{87},
 \eea
 where we divided the total energy by the mean one-particle energy
 $\langle\omega^g_k\rangle \approx \langle\omega^{(2)}\rangle (a_I)$
defined in Eq.~(\ref{ce-1}).
For numerical estimations we use $\widetilde{\gamma}^{(g)}\approx 0.03$.
The number of the primordial gravitons is compatible with the number of
the CMB photons as it was predicted in Ref.~\cite{Babak:1999dc}.

The main result of this Section consists in the evaluation  of the primordial
graviton number~(\ref{H-2abr}). We suppose that the Casimir energy is defined
by the total ground state energy of created excitations, see Eq.~(\ref{H-2a}).

\section{Interaction with fermions \label{sect_matter}}

In this Section, in order to compare our model with the standard approach based on the
Einstein's equations, we consider the interaction with matter fields.

Let us consider  Einstein's equations
 \be\label{m-1R}
 g^{\mu\lambda} \left[ R^{(4)}_{\lambda\nu}(g)
 -\frac{1}{2}g_{\lambda\nu}R^{(4)}(g) \right]
 = - 3 g^{\mu\lambda} T^{\rm matter}_{\lambda\nu}.
 \ee
Here
 \be\label{m-2R}
 T^{\rm matter}_{\mu\nu}=
 -\frac{\delta W_{\rm matter}[g,F^{(n)})]}{\delta g_{\mu\nu}}
 \ee
 is the matter energy momentum tensor in the units (\ref{units1}).

These  equations  are derived by means of the variation
of the Hilbert action $\delta W_{\rm H}/\delta g_{\mu\nu}=0$, where
 \be\label{m-1}
 W_{\rm H}(g,F^{(n)})=\int d^4x \left[-\sqrt{-g}\frac{R^{(4)}(g)}{6}\right]
 +W_{\rm matter}[g,F^{(n)}].
 \ee
Equations (\ref{m-1R}) for the metric components $g_{00}$ and $g_{0j}$ were treated as
four first class constraints (in the Dirac definitions~\cite{dir}). They are consequences of
the general coordinate transformations $x\to \widetilde{x}=\widetilde{x}({x})$
considered  as diffeomorphisms.

In order to realize the  Weyl's idea of conformal symmetry, Dirac had employed
 the conformal transformations
 \bea\label{mC-1}
 g&=&e^{-2D}\widetilde{g},
 \\ \label{mC-2}
 F^{(n)}&=&e^{nD}\widetilde{F}^{(n)}
 \eea
 in the Hilbert action with the aid of the scalar dilaton $D$. As a result, he revealed
 the hidden conformal symmetry of the GR~\cite{Dirac_73}. The identification of the
 new conformal (widetilde) fields with the observational quantities,
 including the metric components $\widetilde{g}, \widetilde{F}^{(n)}$
 is the basic idea of the conformal cosmology~\cite{Behnke:2001nw,Riess_2001,Banerjee:2000}.

 In order to include fermions, we use the Fock simplex in the tetrade
formalism~\cite{Fock:1929vt}:
 \be\label{m-f1}
 W_{\rm matter}[g,\Psi]=\int d^4x \sqrt{-g}\left[-{\overline{\Psi}}i\gamma_{(\beta)}
 D_{(\beta)}\Psi-m_0\overline{\Psi} \Psi\right],
 \ee
where $\gamma_{(\beta)}=\gamma^\mu e_{(\beta)\mu}$ are the Dirac $\gamma$-matrices,
summed with tetrades $e_{(\beta)\nu}$, and $m_0$ is the present-day fermion mass.
The covariant derivative
\be\label{f1-3}
 D_{(\sigma)}=\partial_{(\sigma)}+\frac{i}{2}[\gamma_{(\alpha)},
 \gamma_{(\beta)}]v_{(\alpha) (\beta),(\sigma)}
 \ee
is given by Eqs. (\ref{fock}) and (\ref{carta1}).

Next, we use the Dirac-ADM foliation $(4\to 3+1)$ of the tetrades with the lapse function and
the shift vector~\cite{dir} given in Section~2.
The Dirac's Hamiltonian approach to the theory begins from the determination of the
first class primary constraints. They mean the zero momenta of the time metric components $N, N^j$.
The first class primary constraints lead to the first class secondary constraints
 \bea \label{2c-a}
 && P_{N} = \frac{\partial {\cal L}_{\rm H}}{\partial (\partial_0 { N})} = 0
\quad \Rightarrow \quad \frac{\delta W_{\rm H}}{\delta { N}} = 0,
  \\ \label{2c-b}
 && P_{{N}^j} = \frac{\partial {\cal L}_{\rm H}}{\partial (\partial_0 { N}^j)} = 0
 \quad \Rightarrow \quad \frac{\delta W_{\rm H}}{\delta { N}^j} = 0,
 \eea
where ${\cal L}_{\rm H}$ is the Lagrangian of the Hilbert action $W_{\rm H}=\int d^4x {\cal L}_{\rm H}$.
The first class secondary constraints are supplemented by the second class constraints~(\ref{h-2g})
and~(\ref{D-1}) related to gauge fixing.

The relations between the Conformal and the Standard models can be illustrated using the mass part of the fermion
action
\be\label{mass-1}
 W_{\rm m}[g,\Psi]=-\int d^4x \sqrt{-g}m_0\overline{\Psi} \Psi,
\ee
and the set of its transformations into conformal variables:
\bea
&& g_{\mu\nu}=e^{-2D}\widetilde{g}_{\mu\nu},\qquad \Psi = e^{3D/2}\widetilde\Psi,
\\ \nonumber
&& \sqrt{-\widetilde{g}}=\sqrt{\widetilde{g}_{00}}=e^{-2D}N,\qquad |g_{ij}^{(3)}|=e^{-3D}.
\eea
As a result, we obtain
 \be \label{m-1m}
 W_{\rm m}[N,\widetilde{\Psi},D]=-\int d^4x \sqrt{-\widetilde{g}}\;
 m_0\overline{\widetilde{\Psi}} \widetilde{\Psi}e^{-D}=-\int d^4x N\;
 m_0\overline{\widetilde{\Psi}} \widetilde{\Psi}e^{-3D}.
 \ee
One can see that the variations of the action with respect to  $N$ and $D$
 \bea \label{N-1a}
 N\frac{\delta W_{\rm m}[N,\widetilde{\Psi},D]}{\delta N}, 
 \qquad
\frac{\delta W_{\rm m}[N,\widetilde{\Psi},D]}{\delta D} 
 \eea
is nothing else but a linear combination of the Einstein's equations~(\ref{m-1R}),
{\it i.e.} variations of the action~(\ref{m-f1}) in $g$.
Thus, the classical tests of general relativity including:
perihelion precession of Mercury,
deflection of light by the Sun,
gravitational redshift of light, and
gravitational lensing
are completely fulfilled in our case.
This correspondence between the GR and its dilatonic version
was already discussed by Dirac~\cite{Dirac_73}.
Obviously, separation of the dilaton field into zeroth and non-zeroth harmonics
suggested in our approach does not change local gravitational interactions
with matter, since in the interactions we have always the whole $D=\langle D\rangle+\overline{D}$.

\section{Vacuum creation of Higgs bosons \label{sect_Higgs}}

In our model the interactions of scalar bosons and gravitons with the dilaton
can be treated on the same footing~\cite{Barbashov:2005hu}.
Using this fact, we would like to consider
the intensive creation of the Higgs scalar particles from the vacuum.

To proceed we have to add the SM sector to the theory under construction.
In order to preserve the common origin of the conformal symmetry breaking by the
Casimir vacuum energy, we have to exclude the unique
dimensionful parameter from the SM Lagrangian, {\it i.e.} the Higgs
term with a negative squared tachyon mass.
However, following Kirzhnits~\cite{linde1}, we can include the vacuum
expectation of the Higgs field $\phi_0$, so that:
$\phi=\phi_0+\overline{h}/[a\sqrt{2}],~\int d^3x \overline{h}=0$.
The origin of  this vacuum expectation value $\phi_0$ can be associated with
the Casimir energy arising as a certain external initial data at $a=a_{\rm Pl}$.
In fact, let us apply the Plank least action postulate to the Standard Model action:
 \bea
 W_{\rm SM}(a_{\rm Pl}) \sim \lambda_{\rm SM}\,\, \phi_0^4\,\, a^4_{\rm Pl}
 V_{\rm hor}^{(4)}(a_{\rm Pl}) = 2\pi,
 \eea
where $\lambda_{\rm SM}\sim 1$ is the Higgs self-coupling and $V_{\rm hor}^{(4)}(a_{\rm Pl})$ is given by
Eq.(\ref{Pl-1}). The relation gives
$\phi_0\approx a_{\rm Pl}^3 H_0$, in agreement with its value in Table~1.

 The standard vacuum stability conditions at $a=1$
 \bea
<0|0>|_{\phi=\phi_0}=1, \qquad \frac{d<0|0>}{d\phi}\bigg|_{\phi= \phi_0}=0
 \eea
yield the following constraints on the Coleman--Weinberg effective potential
of the Higgs field:
 \bea
 V_{\rm eff}(\phi_0)=0, \qquad \frac{dV_{\rm eff}(\phi_0)}{d\phi_0}=0.
 \eea
It results in a zero contribution of the Higgs field vacuum
expectation into the Universe energy density. In other words, the SM
mechanism of a mass generation can be completely repeated in the framework
of our approach to the spontaneous symmetry breaking.

In particular, one obtains that the Higgs boson mass is determined from
the equation $V''_{\rm eff}(\langle \phi\rangle)=M_h^2$. Note that
in our construction the Universe evolution is provided by the
dilaton, without making use of any special potential and/or any
inflaton field. In this case we have no reason to spoil the
renormalizablity of the SM by introducing the non-minimal
interaction between the Higgs boson and the gravity~\cite{Bezrukov:2007ep}.

In the approximation (\ref{mfa-11}) of theory (\ref{1-3n})
supplemented by the  Standard Model the Higgs bosons are
described by the action
 \bea
 \label{hh-2s}
 {W}_h&=& \int\limits_{}^{}d\tau
 \sum\limits_{\textbf{k}^2\not =0}
 \frac{v^h_{\textbf{k}}v^h_{-\textbf{k}}\!-\!
 {h}_{\textbf{k}}{h}_{-\textbf{k}}a^2{\omega^h_{0{\bf k}}}^2}{2}
 =\sum\limits_{\textbf{k}^2\not =0}
 p_{-\textbf{k}}^hv^h_{\textbf{k}}-\textsf{H}^h_{\tau},
 \eea
where
 \bea\label{2-10hh}
 \omega^h_{0{\bf k}}({a})=\sqrt{{\bf k}^2+a^2M^2_{0\rm h}}
 \eea
is the massive one--particle energy with respect to the conformal
time interval.

There are values of the scale factor $a$,  when the mass term in the
one--particle energy  is
less than the conformal Hubble parameter value $a M_{0\rm h}<H_0a^{-2}$.
As a result, the Casimir energy for the Higgs particles  coincides
with the graviton one at the considered epoch:
 \bea\nonumber \textsf{H}^h_{\rm Cas}&\simeq& \sum\limits_{\bf
 k}\frac{{\sqrt{{\bf k^2}}}}{2}=\textsf{H}^g_{\rm Cas}.
 \eea

In this case the
calculation of the scalar particle creation energy completely  repeats
the scheme for the graviton creation, discussed above.

Assuming thermalization in the primordial epoch, we expect that the occupation number
of the primordial Higgs bosons is of the order of the known CMB photon one
 \be\label{nu-1}
 \textsf{N}_h \sim \textsf{N}_\gamma=411 {\rm cm}^{-3}\cdot
 \frac{4\pi r_h^3}{3}\simeq 10^{87}.
 \ee
 We point out that this number is of the order
of (\ref{H-2abr}).  Thus, the CGR provides
a finite occupation number of the produced primordial particles. Note
that in other approaches~\cite{grib80} a subtraction is used
to achieve a finite result.
Moreover, the number of produced particles happens to be of the
order of the known CMB photon number. To our opinion this coincidence
supports our model, since the number of photons can naturally inherit
the number of primordial Higgs bosons (if one considers the photons
as one of the final decay products of the bosons).
According to our model, the relativistic matter has been created very soon after
the Planck epoch at $z_{\rm Pl}\simeq 10^{15}$. Later on it cooled down and
at $z_{\rm CMB}\simeq 1000$ the CMB photons decoupled from recombined ions
and electrons as discussed by Gamow. In our model the CMB
temperature is defined directly from the Hubble parameter
and the Planck mass (related to the Universe age $a_{\rm Pl}$).
It is a result of the continuous cooling of the primordial relativistic
matter till the present day described by Eq.~(\ref{ce-1}).

Note that the obtained occupation number~(\ref{nu-1})
corresponds to the thermalized system of photons with the mean wave length
$\lambda_{\rm CMB}$ (at the temperature $T\simeq 3^\circ$~K)
in the finite volume $V_0\sim H_0^{-3}$:
 \be\label{U-1}
 \left(\textsf{N}_\gamma\right)^{1/3} \simeq 10^{29} \simeq \lambda_{\rm CMB} H_0^{-1}.
 \ee

As concerns vacuum creation of spinor and vector SM particles, it is
known~\cite{grib80} to be suppressed very much with respect to the one of scalars
and gravitons.

\section{Summary
\label{sect_Summ}}

We developed a Hamiltonian approach to the gravitational model, formulated
as the nonlinear realization of joint affine and conformal symmetries proposed
long ago in~\cite{Deser:1970hs,Dirac_73,og-73,og-74}. With the aid of the
Dirac-ADM foliation, the conformal and affine symmetries provide
a natural separation of the dilaton
and gravitational dynamics in terms of the  Maurer-Cartan forms.
As the result, the exact solution, Eqs.~(\ref{CC-2}) --- (\ref{Dirac-c2}), of the energy constraint
yields the diffeo-invariant evolution operator in the field space.

In the CGR, the conformal symmetry breaking happens due to the
Casimir vacuum energy (\ref{cas-1})--(\ref{cas-3}). This energy is
obtained as a result of the quantization scheme of the Hamiltonian dynamics
proposed in Sec.~3. In our approach, the Casimir vacuum energy
provides a good description of SNe Ia data~\cite{SN2}
in the conformal cosmology~\cite{Behnke:2001nw,Zakharov:2010nf}\footnote{In these
papers the rigid state was associated with a homogeneous kinetic energy of a free
scalar field.}. The diffeo-invariant dynamics in terms of the Maurer-Cartan
forms with application of the affine symmetry condition leads to the reduction
of the  graviton representation to the  one-component field.
The affine graviton strong wave yields
the effect of expansion (or contraction) in the hypersurface perpendicular to the
direction of the wave propagation.
We demonstrated that the Planck least action postulate applied to the
Universe limited by its horizon provides the value of the cosmological
scale factor at the Planck epoch. A hierarchy of cosmological energy
scales for the states with different conformal weights is found.
The intensive creation of primordial gravitons and
Higgs bosons is described assuming that the Casimir vacuum energy is the source
of this process.
We have calculated the total energy of the created particles, Eq.~(\ref{H-2a}), and
their occupation numbers, Eqs.~(\ref{H-2abr}) and (\ref{nu-1}).

The presented model is under development. To completely establish or discard it, one
has to consider various other problems, including the quantization of the
gravitational field, the CMB power spectrum
anisotropy, baryon asymmetry, thermalization of primordial particles {\it
etc.} Evidently, these problems require a dedicated studies and are left
for the future.

\begin{acknowledgements}
The authors would like thank M.~Bordag, S.~Deser, D.~Ebert, A.~Efremov, V.~Gershun,
Yu.~Ignatev, E.~Lukierski, and A.~Zheltukhin for useful discussions. ABA is grateful to the Dynasty foundation. 
VNP and AB were supported in part by the Bogoliubov-Infeld program.   
AFZ is grateful to the JINR Directorate for a support.
\end{acknowledgements}

\appendix
\section{Dirac Hamiltonian Dynamics  in  Terms
of the Maurer--Cartan Forms \label{sect_AppA}}
\renewcommand{\theequation}{A.\arabic{equation}}
\setcounter{equation}{0}
For the sake of comparison of our approach with the standard Dirac one we
reformulate the latter in terms of the Maurer-Cartan forms. In order have a more general
consideration, we include in the action an electromagnetic field
$F_{\mu\nu}=\partial_\mu A_\nu -\partial_\nu A_\mu$ and a scalar field $Q$
 \bea \label{1-1dc}
 \widetilde{W}[g,A,Q] &=& -\int d^4x {\sqrt{-g}}\;
 \Bigg{( }\frac{1}{6} R^{(4)}(g)
 \quad - \frac{1}{4}F_{\mu\alpha}F_{\nu\beta}g^{\mu\nu}g^{\alpha\beta}
 \nonumber \\
 &+&\partial_\mu Q\partial_\nu Q g^{\mu\nu}\Bigg{)}.
 \eea
Remind that we use the natural units
 \be\label{1-1u}
 \hbar=c=M_{\rm Planck}\sqrt{{3}/({8\pi}})=1.
 \ee

With the aid of the definition of the tetrade components
Eqs.~(\ref{Gold-1}), we obtain the action (\ref{1-1dc})
 \bea
 \label{1-3dgl}
 \widetilde{W}&=&\int\limits_{}^{}
 d^4x{N}\left[{\cal L}_D+{\cal L}_g+{\cal L}_A+{\cal L}_Q\right].
 \eea
Here, the Lagrangian densities are
 \bea\nonumber
{\cal L}_D&=&-{v^2_{{D}}}-\frac{4}{3}e^{-7D/2}\triangle e^{-D/2},
 \\\nonumber
{\cal L}_g&=&\frac{1}{6}\left[v_{(a)(b)}v_{(a)(b)}-e^{-4D}R^{(3)}({\bf e})\right],
 \\\nonumber
{\cal L}_A&=&\frac{1}{2}\left[{e^{2 D}v^2_{(b)(\rm A)}}-{e^{-2D}}F_{ij}F^{ij}\right],
 \\\label{B-4}
{\cal L}_Q&=&e^{2D}{(v_{{Q}}+v_D{\widetilde{Q}})^2}
 -e^{-2D}\left(\partial_{(b)}{\widetilde{Q}}+\partial_{(b)}
 D\widetilde{Q}
\right)^2;
 \eea
and
 \bea
 \nonumber
 v_{Q}&=&\frac{1}{N}\left[(\partial_0-N^l\partial_l){\widetilde{Q}}
+\partial_lN^l/3\right],\\\label{1-3-d}
 v_{{D}}&=&\frac{1}{N}\left[(\partial_0-N^l\partial_l){D}
+\partial_lN^l/3\right],\\\label{1-3-dD}
 \label{1-3-dvvc}
 v_{(a)(b)}&=&\frac{1}{N}\biggl[\omega^R_{(a)(b)}(\partial_0-N^l\partial_l)+
 \partial_{(a)}N^{\bot}_{(b)}
 +\partial_{(b)}N^{\bot}_{(b)} \biggr],
 \\\nonumber
 \label{1-28}
 v_{(b)(\rm A)}&=&\frac{1}{N}{\bf e}^i_{(a)}
 \left[\partial_0A_i-\partial_iA_0+F_{ij}N^j\right]
 \eea
 are velocities of the metric components and fields,
 $\triangle =\partial_i[{\bf e}^i_{(a)}{\bf e}^j_{(a)}\partial_j]$ is
the  Beltrami-Laplace operator,
and $R^{(3)}({\bf{e}})$ is a three-dimensional spatial
curvature expressed in terms of triads ${\bf e}_{(a)i}$ (for the sake of discussion
we use $\widetilde{{\bf e}}\rightarrow {\bf e}$ in Appendix
A),
 \bea \label{1-17a}
 R^{(3)}&=&R^{(3)}({\bf e})-
 \frac{4}{3}{e^{7D/2}}\triangle e^{-D/2},
 \\\label{1-17}
 R^{(3)}({\bf e})&=&-2\partial^{\phantom{f}}_{i}
 [{\bf e}_{(b)}^{i}\sigma_{{(c)|(b)(c)}}]-
 \sigma_{(c)|(b)(c)}\sigma_{(a)|(b)(a)}
 \nonumber \\
 &+& \sigma_{(c)|(d)(f)}^{\phantom{(f)}}
 \sigma^{\phantom{(f)}}_{(f)|(d)(c)},
 \\ \label{1-18}
 \sigma_{{(c)}|(a)(b)} &=& [\omega^L_{(a)(b)}(\partial_{(c)})+
 \omega^R_{(a)(c)}(\partial_{(b))} -
 \omega^R_{(b)(c)}(\partial_{(a)})],
 \nonumber \\ \label{1-19}
 \omega^R_{(a)(b)}(\partial_{(c)})&=&\frac{1}{2}\left[
 {\bf e}^j_{(a)}\partial_{(c)}{\bf e}^j_{(b)}
 +{\bf e}^i_{(b)}\partial_{(c)}{\bf e}^i_{(a)}\right],
 \\ \label{1-20}
 \omega^L_{(a)(b)}(\partial_{(c)})&=&\frac{1}{2}\left[
 {\bf e}^j_{(a)}\partial_{(c)}{\bf e}^j_{(b)}
 -{\bf e}^i_{(b)}\partial_{(c)}{\bf e}^i_{(a)}\right],
 \eea
where $\triangle =\partial_i[{\bf e}^i_{(a)}{\bf
e}^j_{(a)}\partial_j]$ is the Beltrami-Laplace operator.

With help of the Legendre  transformations $v^2/N=pv-Np^2/4$ we determine
momenta
 \bea \label{1-17g}
 P_{(a)(b)}&=&\frac{v_{(a)(b)}}{3},
 \\
 \label{1-17D}
 P_{D}&=& 2{v_{{D}}},\\ \nonumber
 P_{Q}&=&2{v_{{Q}}},\\
 \nonumber
 P_{\rm A(b)}&=&{v_{\rm A(b)}}.
 \eea
Consequently, the total action  (\ref{1-3dgl}) is
 \bea\label{1-3dgc}
  \widetilde{W}&=&\!\! \int\limits_{ }^{ }\!\! d^4x \biggl[
 P_Q\left(\partial_0{{\widetilde{Q}}}+\partial_0{{D}}\widetilde{Q}\right)
 + P_{(a)(b)}\omega_{(a)(b)}^R(\partial_0)
 \nonumber \\
 &+&P_{A{(b)}}\partial_0A_{(b)} -P_D\partial_0{{D}}
 -{\cal C}\biggr],
 \\  \label{1-2c}
 {\cal C} &=&  N{\cal H}+N_{(b)}T_{(b)}
 +A_{(0)}\partial_{(b)}P_{A{(b)}}
 + \lambda_{(0)}P_D 
 \nonumber \\
 &+& \lambda_{(b)}\partial_k {\bf e}^k_{(b)}
 + \lambda_{A}\partial_{(b)} A_{(b)},
 \eea
where
$N$, $N_{(b)}$ and $A_{(0)}\partial_{(b)}$ with $\partial_{(b)}A_{(b)}=0$
are the Lagrange multipliers of the first
class constraints, $\lambda_{(0)}$, $\lambda_{(b)}$
and $\lambda_{A}$ are the multipliers for the second class ones;
 \bea \label{CC-3}
 {\cal H}& =&-\frac{\delta \widetilde{W}}{\delta N}={\cal H}_D+{\cal H}_g
 +{\cal H}_A+{\cal H}_Q,
 \\
 \label{CC-4a}
 {\cal H}_D&=&-\frac{P^2_{D}}{4}-\frac{4}{3}e^{-7D/2}\triangle e^{-D/2},\\\label{3g}
 {\cal H}_g&=&\left[6{P^2_{(a)(b)}}+
 \frac{e^{-4D}}{6}R^{(3)}({\bf e})\right],\\\label{3A}
 {\cal H}_A&=&
 \frac{e^{-2 D}}{2}\left[{P_{i(\rm A)}P^i_{(\rm A)}}+
F_{ij}F^{ij}\right],\\\label{3Q}
 {\cal H}_Q&=&\!e^{-2D}\!\left[\!\frac{P_{{Q}}^2}{4}
 \!+\! \left(\!\partial_{(b)}{Q}\!+\!\partial_{(b)}
 D{Q}\right)^2\right],\\\label{3T}
T_{(0)(a)}\!& = &\!-\!{\bf e}_{(b)}^i\frac{\delta W} {\delta
{N}_{i}}\!=\!
  -\partial_{(b)}P_{(b)(a)}\!+\!\widetilde{T}_{(0)(a)},
 \eea
and
  $\widetilde{T}_{(0)(a)} =
\sum\limits_{F=\overline{\phi},\overline{Q},\widetilde{F}}^{}P_F\partial_{(a)}F$
   are the   energy-momentum tensor
components. Dirac added the second class (gauge) constraints~\cite{dir}:
 \bea
 \label{h-2g}
 && \partial_k {\bf e}^k_{(b)} = 0,\\
 \label{D-1}
 && P_{{D}} =  0 ~~~\to~~~  \partial_0 (e^{-3D})
 +\partial_l (N^le^{-3D})=0.
 \eea
The first three of them fix  spatial coordinates~\cite{dir},
 and   ${P_{{D}}}=0$ is known as the minimal surface  constraint~\cite{Faddeev:1973zb}
distinguished by the co-moving frame of reference. Using the decomposition
 \bea
 \label{dv-7}
 N_{(b)}&=&N^{|\!|}_{(b)}+N^{\bot}_{(b)},\\
 \label{dv-8}
 \partial_{(b)}N^{|\!|}_{(b)}&=& \partial_jN^j,\\
 \label{dv-9}
 \partial_{(b)}N^{\bot}_{(b)}&=&0,
 \\ \label{c-6}
 P_{(b)(a)} &=& P^{\bot}_{(b)(a)}+\partial_{(a)}
 f^{\bot}_{(b)}+\partial_{(b)} f^{\bot}_{(a)}
 \eea
and the solution of the constraint (\ref{3T}),
one can represent the squared momentum in Eq. (\ref{3g}) as
 \be \label{c-7}
 P^2_{(b)(a)}=(P^{\bot}_{(a)(b)})^2+[\partial_{(a)}
 f^{\bot}_{(b)}+\partial_{(b)} f^{\bot}_{(a)}]^2,
 \ee
where $f^{\bot}_{(a)}$ satisfies the equations
 \be \label{c-8}
 \triangle f^{\bot}_{(a)}=\widetilde{T}_{(0)(a)}.
 \ee
The second class constraint (\ref{D-1}) leads to one more secondary constraint
 ${\delta W}/{\delta D}=-T_D=0$, namely,
 \bea\label{e2t}
 && (\partial_\tau-{\cal N}_{(b)}\partial_{(b)})P_{{D}} = T_D,
 \\ \label{e2tc}
 && T_D=\frac{4}{3}\left[7{\cal N}e^{-7D/2}\triangle e^{-D/2}
 +  e^{-D/2}\triangle[{\cal N}e^{-7D/2}]\right]
 \nonumber \\ && \qquad
 - \partial_D[{\cal H}_g+{\cal H}_A+{\cal H}_Q].
 \eea
In Ref.~\cite{Faddeev:1973zb} the Hamiltonian approach to GR is defined
in the class of functions $g_{\mu\nu}(x^0,{\bf x})=\eta_{\mu\nu}+O(1/|\textbf{x}|)$, where
$\eta_{\mu\nu}={\mathrm{Diag}:}(1,-1,-1,-1)$. As a
result, such a theory excludes cosmological evolution.

However, beginning with the pioneering Friedmann results~\cite{Friedman:1922kd} and
continuing with the modern development~\cite{Giovannini,lif,Mukhanov:1990me},
the cosmological evolution can
be incorporated into the gravitational theory with non-flat space-time within the infrared dynamics of
the type of the zeroth mode sector $g_{\mu\nu}(x^0)\not =\eta_{\mu\nu}$ (see
Appendix C). In the paper  we  follow this direction.

\section{Dilaton Cosmological Perturbation Theory
\label{sect_AppB}}
\renewcommand{\theequation}{B.\arabic{equation}}
\setcounter{equation}{0}

 Recall that in general case the local energy density (\ref{CC-3}) is
 \bea\label{CC-1a}
  \widetilde{{\cal H}}=
  -\frac{4}{3}{e}^{-7D/2}
 \triangle  {e}^{-D/2}+ \!\!\!\!
  \sum\limits_{J=0,2,3,4,6} {e}^{-JD}{\cal T}_J(\widetilde{F}),
\eea
 where $\triangle =\partial_i[{\bf e}^i_{(b)}{\bf e}^j_{(b)}\partial_j]$
is the  Beltrami-Laplace operator. The sum is over of the densities of states: rigid
$(J=0)$, radiation $(J=2)$, matter $(J=3)$, curvature $(J=4)$, $\Lambda$-type term
$(J=6)$, respectively,
 in terms of the conformal fields
\be\label{ratio-1b}
 \widetilde{F}^{(n)}=e^{nD}{F}^{(n)},\ee
where  $(n)$ is the conformal weight.

In this case, the equation of the nonzeroth harmonics (\ref{e2tc}) takes the
form~\cite{Barbashov:2005hu}
 \bea \label{CC-4}
  T_D-{\langle T_D\rangle}=0,
 \eea
where
 \bea \label{1-D}
 {T}_{D} &=& \frac{2}{3}
 \left\{7{\cal N}{e}^{-7D/2}
 \triangle {e}^{-D/2}+{e}^{-D/2} \triangle
 \left[{\cal N}{e}^{-7D/2}\right]\right\}
 \nonumber \\
 &+& {\cal N}\sum\limits_{J=0,2,3,4,6}J {e}^{-JD}{\cal T}_J.
 \eea
One can solve all Hamiltonian equations (\ref{CC-1}), (\ref{CC-1a}), and (\ref{CC-4})
to define simplex components
 \bea \label{dg-2aab}
 &&{\widetilde{\omega}}_{(0)}=e^{-2{D}}{\cal N} d\tau,~~~
{\cal N}=\frac{\langle\sqrt{\widetilde{{\cal
H}}}\rangle}{\sqrt{\widetilde{{\cal H}}}},
\\
 \label{dg-3ab}
&&{\widetilde{\omega}}_{(b)}=dX_{(b)}-X_{(c)}\omega^R_{(c)(b)}
+{\cal N}_{(b)}d\tau.
 \eea
Recall that in the lowest order of perturbation theory with respect to the
Newton-type potential $\omega^R_{(c)(b)}$ describes the free one-component
transverse strong gravitational wave considered in Section~3. The
longitudinal  component of the shift vector ${\cal N}_{(b)}$ is
unambiguously determined by the  constraint (\ref{D-2a}) that becomes $\partial_\eta
e^{-3\overline{D}}+
 \partial_{(b)}\left(e^{-3\overline{D}}{\cal N}_{(b)}\right)=0$.

For the small deviations ${\cal N}e^{-7\overline{D}/2}=1-{\nu}_1$ and
 $e^{-\overline{D}/2}=1+{\mu}_1+\ldots$ the
 first orders of Eqs.  (\ref{CC-1a}) and (\ref{1-D}) take the form
 \bea \label{e1-2} \nonumber
 &&[-\hat{\triangle}+14\rho_{(0)}-\rho_{(1)}]\mu_{1} +
   2\rho_{(0)}\nu_1=\overline{\cal T}_{(0)},
 \\ \label{ec1-2}
 &&[7\cdot 14\rho_{(0)}\!-\!14\rho_{(1)}+\rho_{(2)}]\mu_1
 +[-\hat{\triangle}+
14\rho_{(0)}-\rho_{(1)}]\nu_1 =
7\overline{\cal T}_{(0)}-\overline{\cal T}_{(1)},
 \eea
 where
 \bea\label{ec1-3}
 && \rho_{(n)}=\langle{\cal T}_{(n)}\rangle \equiv \!\!\!\!
 \sum\limits_{J=0,2,3,4,6} \!\!\! (2J)^n(1+z)^{2-J}
  \langle{\cal T}_{J}\rangle , \\
\label{ec1-4}
&& {\cal T}_{(n)}= \sum\limits_{J=0,2,3,4,6}(2J)^n(1+z)^{2-J}
{\cal T}_{J}.
 \eea

In the first order of perturbation with respect to the Newton coupling
constant, the lapse function and the dilaton take the forms (see also
\cite{Barbashov:2005hu})
  \bea \label{12-17}
 e^{-\overline{D}/2}
 &=&1+\frac{1}{2}\int d^3y\Bigg{[}G_{(+)}(x,y)
\overline{T}_{(+)}^{(D)}(y)+G_{(-)}(x,y) \overline{T}^{(D)}_{(-)}(y)\Bigg{]},
 \\ \label{12-18}
 {\cal N}e^{-7\overline{D}/2}
 &=&1-\frac{1}{2}\int d^3y\Bigg{[}G_{(+)}(x,y)
 \overline{T}^{(N)}_{(+)}(y)+G_{(-)}(x,y) \overline{T}^{(N)}_{(-)}(y)\Bigg{]},
  \eea
where $G_{(\pm)}(x,y)$ are the Green functions satisfying the equations
 \be
 \label{2-19}\nonumber
 [\pm  m^2_{(\pm)}- \triangle ]G_{(\pm)}(x,y)=\delta^3(x-y).
 \ee
 Here
 \bea
 \label{1cur1}
\nonumber 
 m^2_{(\pm)} &=& H_0^2 \frac{3(1\!+\!z)^2}{4}\Bigg{[}{14(\beta\pm 1)}
\Omega_{(0)}(a)\mp \Omega_{(1)}(a)\Bigg{]},
\\ \nonumber 
 \beta &=& \sqrt{1+[\Omega_{(2)}(a)-14\Omega_{(1)}(a)]/[98\Omega_{(0)}(a)]},
\\
 \overline{T}^{(D)}_{(\pm)} &=& \overline{{\cal T}}_{(0)}\mp7\beta
  [7\overline{{\cal T}}_{(0)}-\overline{{\cal T}}_{(1)}],
 \\ \label{1curv2}
 \overline{T}^{(N)}_{(\pm)} &=& [7\overline{\cal T}_{(0)}
 -\overline{\cal T}_{(1)}]
 \pm(14\beta)^{-1}\overline{\cal T}_{(0)},
 \eea
 are the local currents, and
 \bea
 \label{1cur3}
 \Omega_{(n)}(a)=\sum\limits_{J=0,2,3,4,6}(2J)^n(1+z)^{2-J}\Omega_{J}, \eea
 $\Omega_{J=0,2,3,4,6}= \langle{\cal T}_J\rangle/H_0^2$ are partial
densities of states: rigid, radiation, matter, curvature, $\Lambda$-term,
respectively; $\Omega_{(0)}(a=1)=1$, $1+z=a^{-1}$ and $H_0$ is the Hubble
parameter.

In the context of  these definitions, a full  family of solutions  (\ref{12-17}),
(\ref{12-18}) for the lapse function and the nonzeroth dilaton harmonics
of the Hamiltonian constraints (\ref{h-c1d})-(\ref{Dirac-c2}),
yield a Newton-type potential.
In particular, for a point mass distribution in
a finite volume which corresponds to the nonzero terms with
a)$J=0,3$ in Eq.(\ref{ec1-3}); b)$J=3$ in Eq.(\ref{ec1-4});
c)$J=0,3$ in Eq.(\ref{1cur3}) (otherwise zero), - we have
 \be \label{T-1}
 \overline{{\cal T}}_{(0)}(x)=\frac{\overline{{\cal T}}_{(1)}(x)}{6}
 \equiv \frac{3}{4a^2} M\left[\delta^3(x-y)-\frac{1}{V_0}\right].
 \ee
As a result, solutions (\ref{12-17}) and (\ref{12-18}) are transformed to the
Schwarzschild-type form
 \bea
 \label{N-1}
  e^{-\overline{D}/2}&=&1+
  \frac{r_{g}}{4r}\Bigg{[}\frac{1+7\beta}{2}e^{-m_{(+)}(a)r}
 + \frac{1-7\beta}{2}\cos{m_{(-)}(a) r}\Bigg{]},\\
 \label{N-2}
 {\cal N}e^{-7\overline{D}/2}&=&1-
 \frac{r_{g}}{4r}\Bigg{[}\frac{14\beta+1}{28\beta}e^{-m_{(+)}(a)
 r}+\frac{14\beta-1}{28\beta}\cos{m_{(-)}(a) r}\Bigg{]},
 \eea
 where $r_{g}=M/M^2_{\rm Pl}$, $\beta=5/7$, $m_{(+)}=3m_{(-)}$,
 and $m_{(-)}=H_0\sqrt{3(1+z)\Omega_{\rm Matter} /2}$.
These solutions describe the Jeans-like spatial oscillations of
 the scalar potentials (\ref{N-1}) and (\ref{N-2}) even for the
 case of zero pressure.

 These spatial oscillations can determine the clustering of matter
  in the recombination epoch, when  the redshift is close to the value
  $z_{\rm recomb.}\simeq 1100$.
  Indeed, if we use for the matter clustering parameter
(that follows from spatial oscillations of the modified Newton law
(\ref{N-1}), (\ref{N-2}))
  the observational value ~\cite{a1}
\begin{equation}\label{cl-1}
r_{\rm clustering} \simeq 130\, {\rm Mpc} \simeq\frac{1}{m_{(-)} }
=\frac{1}{ H_0[\Omega_{\rm Matter} (1+z_{\rm recomb})]^{1/2}}.
 \end{equation}
 one obtains  $\Omega_{\rm Matter}\sim  0.2$.
This estimation is in an agreement with the one,
recently discovered in the quest of the large scale periodicity
distribution (see for details in~\cite{Zakharov:2010nf}).

Constraint  (\ref{D-2a}) yields the shift of the coordinate
origin in the process of the evolution
 \be \label{2-23}
 {\cal N}^i=\left(\frac{x^i}{r}\right)
 \left(\frac{\partial_\eta V}{\partial_r V}\right), \qquad
 V(\eta,r)=\int\limits_{}^{r}d\widetilde{r}\;
 \widetilde{r}^2e^{-3\overline{D}(\eta,\widetilde{r})}.
 \ee
In the limit  $H_0=0$ at $a_0=1$, the solutions (\ref{N-1})
and (\ref{N-2}) coincide with the isotropic Schwarzschild solutions:
 $e^{-\overline{D}/2}=1+r_g/(4r)$,~
 ${\cal N}e^{-7\overline{D}/2}=1-r_g/(4r)$,~
 ${\cal N}_{i}=0$.
Solution~(\ref{N-1}) doubles the angle of the photon beam deflection by the Sun
field. Thus, the CGR  provides also the Newtonian limit in our variables.

\section{Conformal Cosmology
\label{sect_CppC}}
\renewcommand{\theequation}{C.\arabic{equation}}
\setcounter{equation}{0}

The distant supernovae data provide a powerful test for all theoretical
cosmological models in spite of the fact that the correctness of the
hypothesis about SNe Ia as the perfect standard candles is still not
proven~\cite{Panagia:2005hr}. However, the first observational conclusion
about accelerating expansion of the Universe and about the existence of the
non-vanishing $\Lambda$-term was made with the cosmological SNe Ia data.

Among different theories that passed this test, see {\it
e.g.}~\cite{Riess_2001,Zhu:2003sq}, there are conformal cosmological
models~\cite{Behnke:2001nw,Behnke_04,Blaschke:2004by,Barbashov:2005hu,Zakharov:2010nf}
which assume  to explain the long distance SNe Ia by  the long dilaton intervals
of the Dirac version of GR~\cite{dir} considered in the present paper.
 This type of cosmological model
naturally emerges from our approach to the GR, which is based on the conformal
symmetry. In this case the unknown dark energy of $\Lambda$-term
is replaced by the well known Casimir vacuum energy of the empty Universe.

The construction of all observable CC-quantities
is based on the {\it conformal postulate} in accord to which each CC-quantity
$F_c^{(n)}$ with conformal weight  $(n)$ is equal to the SC one, $F_s^{(n)}$,
multiplied by the cosmological scale factor to the power $(-n)$:
 \be\label{c-sg}
 F_c^{(n)}=a^{-n}F_s^{(n)}.
 \ee
 In accord with the conformal
postulate (\ref{c-sg}), the CC-time is greater than the SC one, and all
CC-distances, including the CC-luminosity distance $\ell_c$, are longer than
the SC-ones $\ell_s=a\ell_c$, because all intervals are measured by  clocks of
mass {\it Const}$/a$.

The first attempts to analyze SNe Ia data to evaluate parameters of CC models
were made in~\cite{Behnke:2001nw}, where only 42 high redshift type Ia
SNe~\cite{SN2} point were used. Later a slightly extended sample was
analyzed~\cite{Behnke_04}.
 In spite of a small size of the samples used in previous attempts
 to fit CC model parameters, it was
concluded that if $\Omega_{\rm Rigid}$ is significant with respect to the
critical density (\ref{cr-1a}), CC models could fit SNe Ia observational data with a
reasonable accuracy. After that a possibility to fit observational SNe Ia data
with CC models was seriously discussed by different authors~\cite{Riess_2001,Tegmark:2001zc}
among other alternatives.

In both the cosmological models, the dependence of the scale factor ($a$) on
the conformal time ($\eta$) is given by the Einstein---Friedmann equation
\cite{Friedman:1922kd}
 \bea
 \label{a:Omega}
 \left(\frac{da}{d\eta}\right)^2&=&\rho_\eta
=H_0^{2}\Omega({a}),\\
\Omega({a})&\equiv&\Omega_{\Lambda}a^4+\Omega_{\rm Matter}a+\Omega_{\rm
Radiation}+\Omega_{\rm Rigid}a^{-2},\nonumber
 \eea
 where $\Omega({a})$ is the sum of the
partial densities: $\Lambda$--term--state, matter, radiation,  and rigid,
respectively, normalized by the unit
 density $\Omega\Big{|}_{a=1}=1$; $H_0$ is the present--day value of the
Hubble parameter.
One obtains from Eq.(\ref{a:Omega}) the definition
of the horizon
 \bea
\label{1-3hh1}
d_{\rm hor}({a})=
2r_{\rm hor}(z)=
2\int\limits_{a_I\to 0}^{{a}}
d \overline{a}
{\frac{1}{\sqrt{\rho_\eta(\overline{a})}}}
\eea
  Thus, this distance determines
the diameter of the visible Universe ``sphere''.

The best fit to the Supernova data~\cite{SN2} requires a cosmological constant
$\Omega_{\rm Rigid}=0$, $\Omega_\Lambda=0.7$ and $\Omega_{\rm Matter}=0.3$
 in the $\Lambda$CDM model, where the measurable distance is identified
with the world space interval $R=ar$.

In the conformal
cosmology~\cite{Behnke:2001nw}, measurable time and distance are identified with
the conformal quantities $(r,\eta)$.
Therefore, in the CC framework, we have a possibility to consider the Early
Universe evolution~\cite{Arbuzov:2009bi} using the parameters of the CC dark energy
obtained from
the SNe Ia data~\cite{SN2}. In our CC model, the dark energy is the integral of motion
${\rho_{ I\rm Cas}}\simeq {\rho_{ 0\rm Cas}}$
and has the substantial foundation as experimental fact.

\end{document}